\documentclass[twocolumn]{aastex631}
\let\tablenum\relax
\usepackage{siunitx}
\usepackage{amsmath}
\usepackage{natbib}
\usepackage{multirow}

\begin{document}

\title{The H$\alpha$ broadband photometric reverberation mapping of four Seyfert 1 galaxies}

\email{maqinchun@pku.edu.cn; wuxb@pku.edu.cn}
\author{Qinchun Ma}
\affiliation{Department of Astronomy, School of Physics, Peking University, Beijing 100871, China}
\affiliation{Kavli Institute for Astronomy and Astrophysics, Peking University, Beijing 100871, China}

\author{Xue-Bing Wu}
\affiliation{Department of Astronomy, School of Physics, Peking University, Beijing 100871, China}
\affiliation{Kavli Institute for Astronomy and Astrophysics, Peking University, Beijing 100871, China}

\author{Huapeng Gu}
\affiliation{Department of Astronomy, School of Physics, Peking University, Beijing 100871, China}
\affiliation{Kavli Institute for Astronomy and Astrophysics, Peking University, Beijing 100871, China}

\author{Yuhan Wen}
\affiliation{Department of Astronomy, School of Physics, Peking University, Beijing 100871, China}
\affiliation{Kavli Institute for Astronomy and Astrophysics, Peking University, Beijing 100871, China}

\author{Yuming Fu}
\affiliation{Department of Astronomy, School of Physics, Peking University, Beijing 100871, China}
\affiliation{Kavli Institute for Astronomy and Astrophysics, Peking University, Beijing 100871, China}

\begin{abstract}

Broadband photometric reverberation mapping (PRM) have been investigated for AGNs in recent years, but mostly on accretion disk continuum RM. Due to the small fraction of broad emission lines in the broadband, PRM for emission lines is very challenging. Here we present an ICCF-Cut method for broadband PRM to obtain the H$\alpha$ broad line lag and apply it to four Seyfert 1 galaxies, MCG+08-11-011, NGC 2617, 3C 120 and NGC 5548. All of them have high quality broadband lightcurves with daily/sub-daily cadence, which enables us to extract H$\alpha$ lightcurves from the line band by subtracting the contributions from the continuum and host galaxy. Their extracted H$\alpha$ lightcurves are compared with the lagged continuum band lightcurves, as well as the lagged H$\beta$ lightcurves obtained by spectroscopic RM (SRM) at the same epochs. The consistency of these lightcurves and the comparison with the SRM H$\beta$ lags provide supports to the H$\alpha$ lags of these AGNs, in a range from 9 to 19 days, obtained by the ICCF-Cut, JAVELIN and $\chi^2$ methods. The simulations to evaluate the reliability of H$\alpha$ lags and the comparisons between SRM H$\beta$ and PRM H$\alpha$ lags indicate that the consistency of the ICCF-Cut, JAVELIN and $\chi^2$ results can ensure the reliability of the derived H$\alpha$ lags. These methods may be used to estimate the broad line region sizes and black hole masses of a large sample of AGNs in the large multi-epoch high cadence photometric surveys such as LSST in the future.

\end{abstract}


\section{Introduction}
    
Active Galactic Nuclei (AGNs) are powered by the accretion processes onto the central supermassive black holes (SMBHs)\citep{1995PASP..107..803U}. Surrounding the SMBH is the geometrically thin, optically thick accretion disk which generates the UV/optical continuum emission. The continuum light from the central accretion disk travels across the broad line region (BLR), and produces broad emission lines through the photoionization process. AGNs often show large and aperiodic variabilities in all the wavebands, but the origin of which is still not very clear. Several models including the accretion disk instabilities \citep{1998ApJ...504..671K} and the general Poisson process models \citep{2000ApJ...544..123C} have been proposed to characterize the optical variability of AGNs.
    
Reverberation mapping (RM) \citep{1982ApJ...255..419B,1993PASP..105..247P} exploits the time delay $\tau$ between the lightcurves of the optical continuum and broad emission lines to study the size and structure of the BLR. The average size of the BLR is $R_{\rm BLR}=\tau \cdot c$, where $c$ is the speed of light. This method has been proved powerful to estimate the virial mass of the central supermassive black hole,
    
\begin{equation}
M_{\rm BH}=f \frac{R_{\rm BLR}\cdot{\sigma_v}^2}{G},
\end{equation}
where $\sigma_v$ is the velocity dispersion of the broad emission line which can be estimated from the spectrum, $G$ is the gravitational constant and $f$ is a dimensionless factor in order of unity that depends on the geometry and kinematics of the BLR \citep{1999ApJ...521L..95P,2004ApJ...615..645O,2006MNRAS.373..551L,2010ApJ...716..269W,2011MNRAS.412.2211G}. The factor $f$ can be estimated for nearby AGNs whose $M_{\rm BH}$ values have been obtained by both RM and the correlation between $M_{\rm BH}$ and the stellar velocity dispersion \citep{2006A&A...456...75C,2015ApJ...801...38W}.
    
RM has established an empirical relationship between the BLR size and the AGN continuum luminosity \citep{1996ApJ...471L..75K,2000ApJ...533..631K,2006ApJ...644..133B,2009ApJ...694L.166B}, $R_{\rm BLR}\propto L^\alpha$. The theoretical prediction from the photoionization model gives $\alpha = 0.5$ \citep{1990agn..conf...57N}, which has been examined extensively by many observational campaigns \citep{1991ApJ...370L..61K,1996ApJ...471L..75K,1999ApJ...526..579W,2008ApJ...673..703M,2011nlsg.confE..38V,2015ApJS..216....4S,2017ApJ...849..146G}. However, a large number of accurate measurements of $R_{\rm BLR}$ are required to reduce the scatter of the current $R-L$ relation. Because the sizes of BLRs usually range from a few to several hundred light days and the observed time lags are the product of the rest-frame time lags and the time dilation factor $1+z$, monitoring the variability of AGNs can take months to years. Spectroscopic reverberation mapping (SRM) monitors the spectra of AGNs to get the time delay between the continuum and the broad emission line. However, SRM campaigns are expensive because they need a large amount of observational time from intermediate to large optical telescopes. In addition, extracting accurate lightcurves from spectroscopic observations is often difficult due to the uncertainties in the flux calibration process \citep{2008A&A...486...99S,2011MNRAS.416..225S}.

Photometric reverberation mapping (PRM) employs broad bands to trace the AGN continuum and suitable narrow bands to trace the broad emission lines. With small optical telescopes, PRM can monitor AGNs with high efficiency. For example, \citet{2011A&A...535A..73H} used the PRM with a 15-cm telescope VYSOS-6 to obtain the BLR sizes of PG0003+199 and Ark 120, proving the feasibility of the PRM method. Because the narrow bands contain both the emission lines and the underlying continuum, the contribution from the continuum in the narrow bands must be considered. \citet{2012A&A...545A..84P} computed the synthetic $\rm H\beta$ lightcurve by subtracting a scaled broadband lightcurve from the narrow band lightcurve and measured the BLR size of 3C 120, which is very close to the value obtained with SRM \citep{2012ApJ...755...60G}. In addition to the narrow-band PRM, \citet{2019ApJ...884..103K} used three intermediate band filters to get the time lags between the continuum and the $\rm H\alpha$ emission line of five AGNs, which are consistent with the SRM results. \citet{2016ApJ...818..137J} combined the broad and intermediate bands and detect the $\rm H\alpha$ time lags for 13 AGNs at redshift $0.2<z<0.4$.
    
While the narrow and intermediate band PRM methods can increase the signal-to-noise ratio (S/N) of emission lines, they have to limit the range of AGN redshifts and require special filters. The broadband PRM uses a suitable broadband to trace the strong emission line, allowing it to cover the emission line over a wide range of redshifts. Some previous works \citep{2012ApJ...756...73E,2016ApJ...819..122Z} have investigated several AGNs with the broadband PRM. 
The most important advantage is that the broadband PRM can use the multi-epoch data of large photometric sky surveys such as the Zwicky Transient Facility (ZTF) \citep{2019PASP..131a8003M} and the Legacy Survey of Space and Time (LSST of the Vera C. Rubin Observatory) \citep{2017arXiv170804058L}. Such surveys are much more efficient than the narrow or intermediate band PRM campaigns, which are only applicable to a few targets within a narrow redshift range. 
The multi-epoch data of photometric sky surveys have been widely used for continuum reverberation mapping \citep{2019ApJ...880..126H, 2020ApJS..246...16Y}. They usually use the data of broad bands to calculate the continuum lags between two broad bands directly.
An important issue for the broadband PRM of emission lines is that the emission line only contributes a small fraction of the total flux in the broad band, where the continuum is dominant. Therefore the difference between the lightcurves of the line band and the continuum band is usually too small to calculate the emission line time lag directly. We need some new methods to obtain the emission line time lags for the broadband PRM, and make comparisons with the SRM results to examine the methods and results.

We select 4 Seyfert 1 galaxies MCG +8-11-011, NGC 2617, 3C 120 and NGC 5548, which have been widely studied from continuum RM to SRM in previous research. \citet{2017ApJ...840...97F} and \citet{2018ApJ...854..107F} presented the results of the simultaneous continuum RM and H$\beta$ SRM for MCG +8-11-011 and NGC 2617. NGC 2617 is also known as a changing look AGN \citep{2014ApJ...788...48S} from other SRM campaign \citep{2021ApJ...912...92F} and continuum RM campaign \citep{2021MNRAS.503.4163K}. 3C 120 has been widely studied with SRM \citep{1998ApJ...501...82P,2012ApJ...755...60G,2013ApJ...764...47G,2014A&A...566A.106K,2020MNRAS.497.2910H}. \citet{2018A&A...620A.137R} used the narrow band to obtain the H$\alpha$ narrow band PRM lag for 3C 120, which is much larger than the SRM H$\alpha$ lag at a different time epoch, mainly due to the much higher luminosity in this period. NGC 5548 is one of the best-studied AGNs with many RM campaigns in past decades \citep{1993PASP..105..247P,2002ApJ...581..197P,2007ApJ...662..205B,2009ApJ...704L..80D,2010ApJ...716..993B,2016ApJ...827..118L,2019MNRAS.489.1572L,2021ApJ...907...76H}. The AGN Space Telescope and Optical Reverberation Mapping Project \citep[AGN STORM:][]{2016ApJ...821...56F,2017ApJ...837..131P} has done multiwavelength photometric and simultaneous spectroscopic observations of NGC 5548.

\begin{figure*}[!htb]
    \includegraphics[width=1.0\linewidth]{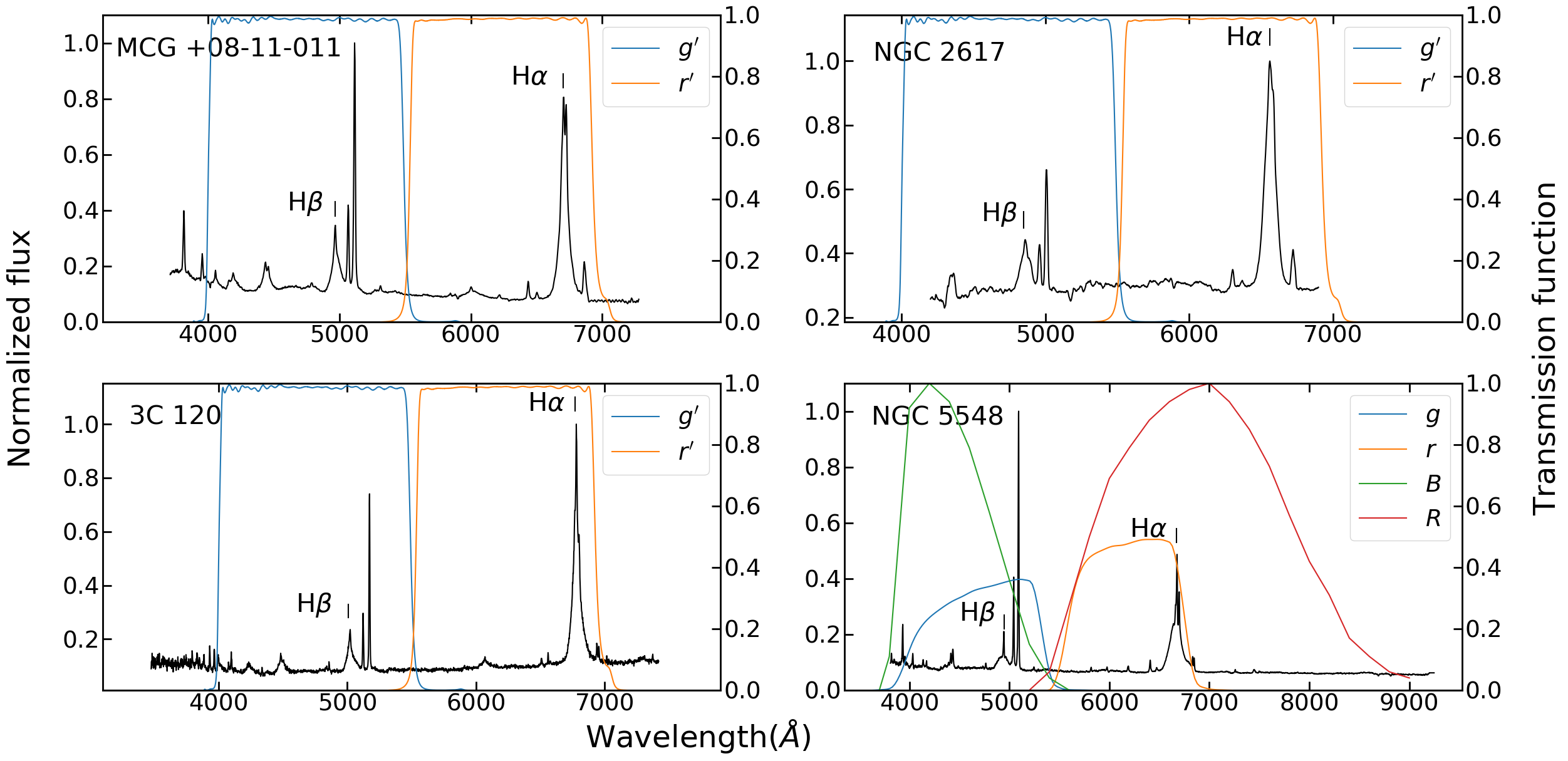}
    \caption{The spectrum of 4 Seyfert 1 galaxies and the transmission functions of the broad bands used for broadband PRM. The $g$, $g'$ and $B$ bands are used for the continuum bands and the $r$, $r'$ and $R$ bands are used to extract the H$\alpha$ line.}
\end{figure*}

This paper is arranged as follows. We describe the target selections in Section 2. The methods to calculate the time lags are presented in Section 3. The calculated H$\alpha$ lag results for 4 Seyfert 1 galaxies are presented in Section 4. We use the simulations and other methods to evaluate the influences of the $\rm H\beta$ and inter-continuum lags in Section 5. A summary is given in Section 6. We adopt the standard $\Lambda{\rm CDM}$ cosmology with $\Omega _m=0.32$, $\Omega_{\rm \Lambda}=0.68$ and $H_{\rm 0}=67\ {\rm km\ s^{-1}\ Mpc^{-1}}$ \citep{2014A&A...571A..16P} in this paper.

\section{Target selections}    

\begin{deluxetable*}{cccccccc}
    \tablecaption{The properties for 4 Seyfert 1 galaxies. \label{tab:mathmode}}
    \tablehead{
    \colhead{Name} & \colhead{Redshift} & \colhead{H$\alpha$ ratio($\%$)} & \colhead{Band} & \colhead{Duration(days)} & \colhead{Epoch} & 
    \colhead{Cadence(days)} & \colhead{$F_{var}$($\%$)}}
    \startdata
    MCG +8-11-011 & 0.0205 & 26 & $r'$ & 93 & 42 & 1.07 & 4.8\\
     &  &  & $g'$ & 156 & 85 & 1.00 & 7.0\\
    NGC 2617 & 0.0142 & 16 & $r'$ & 102 & 127 & 0.62 & 3.7 \\
     &  &  & $g'$ & 145 & 166 & 0.69 & 3.8 \\
    3C 120 & 0.0330 & 23 & $r'$ & 249 & 370 & 1.08 & 6.3 \\
     &  &  & $g'$ & 246 & 392 & 1.06 & 9.1 \\
    NGC 5548 & 0.0172 & 25 & $r$ & 212 & 203 & 0.93 & 3.5 \\
     &  &  & $g$ & 212 & 204 & 0.98 & 6.2 \\
     &  & 19 & $R$ & 226 & 161 & 0.96 & 3.7 \\
     &  &  & $B$ & 225 & 180 & 1.00 & 8.6 \\
    \enddata
\end{deluxetable*}

Broadband PRM requires at least two broadband lightcurves with high photometric accuracy and high observational cadence for AGNs. One band is the continuum band that does not contain strong broad emission lines and another band is the line band with a strong emission line. To examine our methods and results, we select AGNs with both multi-band photometric observations and simultaneous SRM results, so that we can compare the time lags and the shapes of lightcurves of the broadband PRM with the time lags and lightcurves of the SRM. Because the H$\alpha$ line is much stronger than the H$\beta$ line, obtaining the H$\alpha$ time lag is more feasible for broadband PRM. The best way is to use the data of the H$\alpha$ SRM campaigns to compare the broadband PRM H$\alpha$ time lags and lightcurves with the SRM results.
However, the number of H$\alpha$ SRM campaigns is much smaller than H$\beta$ SRM campaigns, and these H$\alpha$ SRM campaigns do not have enough photometric broadband observations or their photometric lightcurves do not have high accuracy and high cadence for broadband PRM. Finally we select four Seyfert 1 galaxies with simultaneous high quality continuum RM and H$\beta$ RM observations. We use the photometric data to obtain the time lags and lightcurves of H$\alpha$ line and compare these results with their SRM H$\beta$ time lags and lightcurves.

Table 1 shows the properties including the photometric durations and the epochs of the line band for 4 Seyfert 1 galaxies. The data of MCG +8-11-011 and NGC 2617 are obtained from \citet{2018ApJ...854..107F} and \citet{2017ApJ...840...97F}. The data of 3C 120 are obtained from \citet{2020MNRAS.497.2910H}. These three targets were observed with the Las Cumbres Observatory \citep[LCO;][]{2013PASP..125.1031B} global robotic telescope network. The data of NGC 5548 are obtained from \citet{2016ApJ...821...56F} and \citet{2017ApJ...837..131P}. Their H$\alpha$ ratios are calculated from the single-epoch spectra shown in Figure 1. The transfer functions of the continuum bands and the line bands used in the broadband PRM for 4 galaxies are also shown in Figure 1. Because the spectrum of the simultaneous H$\beta$ SRM can not cover the whole wavelength range of the line band, we use the single-epoch spectra from other observations and campaigns as the substitutes.  The spectrum of NGC 2617 is obtained from \citet{2021ApJ...912...92F}. The spectrum of 3C 120 is obtained from \citet{2018A&A...620A.137R}. The spectrum of NGC 5548 is obtained from the Sloan Digital Sky Survey \citep[SDSS;][]{2000AJ....120.1579Y}. We obtain the single-epoch spectrum of MCG +8-11-011 using the Beijing Faint Object Spectrograph and Camera (BFOSC) of the Xinglong 2.16-m telescope in China. We use the Grism 4 with a dispersion of $\SI{198}{\angstrom}/{\rm mm}$ and the slit width of $1\arcsec.8$. The spectra are reduced by the standard IRAF routine \citep{1986SPIE..627..733T,1993ASPC...52..173T}. 

We use the LCO $g'$ band as the continuum band and the $r'$ band as the line band for MCG +8-11-011, NGC 2617 and 3C 120. For NGC 5548, besides the SDSS $g$ and $r$ bands, we also use the Johnson/Cousins $B$ and $R$ bands to do the broadband PRM as the comparison. Table 1 and Figure 1 show that all of them have very high photometric cadences and strong H$\alpha$ emission lines for the broadband PRM (we will not distinguish $g'$ and $r'$ from $g$ and $r$ afterwards).

\section{Time Lag Calculations}
\subsection{The ICCF-Cut Method}
\begin{figure*}[htb!]
    \includegraphics[width=1.0\linewidth]{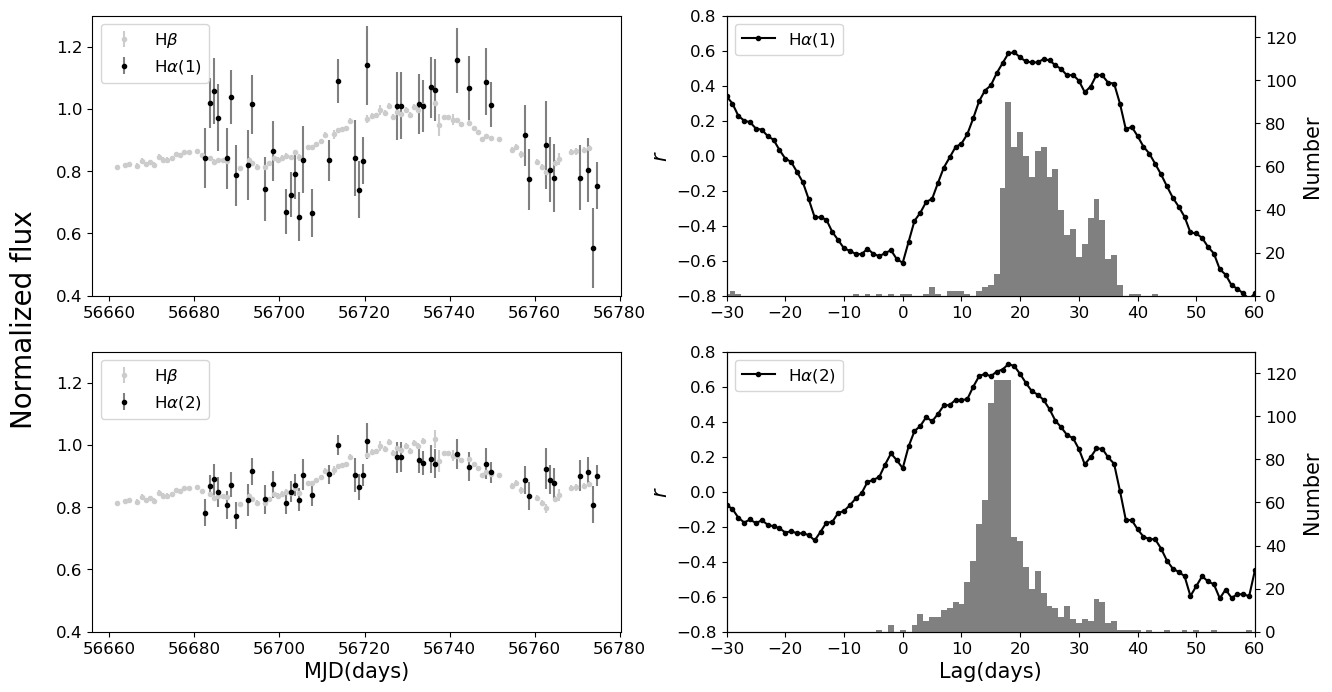}
    \caption{The SRM H$\beta$ and extracted H$\alpha$ lightcurves and the ICCF-Cut lag distributions of MCG +8-11-011 with the $15.7_{-0.5}^{+0.5}$ days SRM ${\rm H}\beta$ lag. The left panel shows the H$\beta$ and extracted ${\rm H}\alpha$ emission lines lightcurves with different methods. The grey points represent the H$\beta$ lightcurves. The labels H$\alpha$(1) and H$\alpha$(2) represent the cases using Eqs. (3) and (4) respectively. All lightcurves have been normalized and the Y-axis has been set to the same scaling for comparison. The right panel shows the ICCF-Cut lag results obtained from the H$\alpha$(1) and H$\alpha$(2). The black lines represent the relations between the cross-correlation coefficient $r$ with the time lag and the grey parts represent the 1000 FR/RSS simulations which represent the lag distributions. The time lag between the $g$ band and the H$\alpha$ line for H$\alpha$(1) and H$\alpha$(2) is $23.8_{-4.8}^{+8.2}$ and $17.3_{-4.3}^{+6.2}$ days respectively.}
\end{figure*}

The interpolated cross-correlation function \citep[ICCF;][]{1986ApJ...305..175G}
calculates the time lag between two different lightcurves directly and has been extensively used in SRM. To get the emission line time lag of the broadband PRM, we need to get rid of the continuum contribution in the line band. First we simply assume that the continuum flux in the line band equals to a fixed fraction $\alpha$ of the flux in the continuum band for each AGN. Under this assumption, we use the line band flux to subtract the continuum band flux with the ratio $\alpha$, and get the lightcurve of ${\rm H}\alpha$ line (hereafter ICCF-Cut),
\begin{equation}
L_{{\rm H}\alpha}(t)=L_{\rm line}(t)-\alpha L_{\rm cont}(t).
\end{equation}
Here $L_{\rm line}(t)$, $L_{\rm cont}(t)$, and $L_{{\rm H}\alpha}(t)$ are the lightcurves of the line band, continuum band, and ${\rm H}\alpha$ emission line respectively. We ignore the influence of ${\rm H}\beta$ variability in the continuum band because the ${\rm H}\beta$ emission line usually contributes less than $5\%$ in the continuum band and is much weaker than the ${\rm H}\alpha$ emission line (which has a $10\sim 30\%$ contribution to the line band). According to the thin disk model, there is a small inter-continuum time lag between the continuum in the $g$ and $r$ bands. Because the inter-continuum lag is usually very small for lower redshift Seyfert 1 galaxies \citep{2016ApJ...821...56F,2018ApJ...854..107F},
we ignore the influence of the inter-continuum lag between the line band and continuum band at first. Further simulation and discussion on the influence of the ${\rm H}\beta$ emission line and the inter-continuum lag will be presented in Section 5. The value of $\alpha$ is associated with the spectral slope of the AGN continuum and usually does not change much within several months. We use the spectral data of the LICK AGN MONITORING PROJECT \citep[LAMP;][]{2010ApJ...716..993B} to examine the variability of $\alpha$ and find that $\alpha$ changes little during the campaign (e.g. $\alpha=1.02\pm 0.07$ for NGC 4748, and $\alpha=1.16\pm 0.09$ for Arp 151).

The $\alpha$ value can be obtained from the transmission functions of the line and continuum band filters and the single-epoch spectrum of an AGN. 
Because the spectra of simultaneous H$\beta$ SRM for these targets can not cover the entire wavelength range of the line band, we use the single-epoch spectra in Figure 1 to calculate $\alpha$:
\begin{equation}
\alpha =(F_{\rm line}-F_{{\rm H}\alpha})/F_{\rm cont}.
\end{equation}
Here $F_{\rm cont}$, $F_{\rm line}$ and $F_{{\rm H}\alpha}$ are the fluxes obtained from the integral of the single-epoch spectrum. However, there are two issues. One is that the flux calibration of the spectrum is usually not as accurate as the photometric flux calibration. Another one is that these spectroscopic and photometric observations have been conducted at different epochs (several years apart from each other), so the spectral index of the AGN continuum and the value of $\alpha$ may change significantly. Here we adopt another method to calculate the value of $\alpha$ for the broadband PRM. We assume the average contribution of the ${\rm H}\alpha$ line in the line band does not change much between the spectroscopic epoch and the mean epoch of the photometric lightcurves. To make sure the extracted ${\rm H}\alpha$ lightcurve containing all the contributions of the ${\rm H}\alpha$ emission line, we use the minimum function:
\begin{equation}
\alpha=\left(1-\frac{F_{{\rm H}\alpha}}{F_{\rm line}}\right)\min\left(\frac{L_{{\rm line},t}}{L_{{\rm cont},t}}\right)
\end{equation}
to calculate the value of $\alpha$. Here $F_{{\rm H}\alpha}$ and $F_{\rm line}$ are measured from the single-epoch spectrum, and $L_{{\rm line},t}$ and $L_{{\rm cont},t}$ are the fluxes at each point obtained from linear interpolated photometric lightcurves. 
For the single-epoch spectrum and two broadband lightcurves, we calculate the $\alpha$ value using Eq. (4) and use it for the observational period.
In this case, adopting the minimum as the $\alpha$ value means that the contribution of ${\rm H}\alpha$ in the line band for each point of the photometric lightcurve will be kept and is larger than the contribution obtained directly from the single-epoch spectrum. However, we can still exclude a large fraction of continuum in the line band while keeping the ${\rm H}\alpha$ contribution as complete as possible. We will discuss the influence of the varying value of $\alpha$ in Section 6.

Here we investigate the difference in the results of applying Eqs. (3) and (4) to obtain the value of $\alpha$. Figure 2 shows an example of the comparison between the results derived from two different $\alpha$ values given by the two approaches above for MCG +8-11-011. By comparing the lightcurves of the extracted H$\alpha$ and the SRM H$\beta$ lines, we find that the extracted H$\alpha$ lightcurve using Eq. (4) is more consistent with the SRM H$\beta$ lightcurve. The value of the cross-correlation coefficient $r$ of using Eq. (4) is higher and the lag distribution is smoother than using Eq. (3). This indicates that the result obtained from Eq. (4) is more reliable. When using Eq. (3), the extracted ${\rm H}\alpha$ lightcurves do not contain enough contribution of the real ${\rm H}\alpha$ line variability to obtain the time lag. This comparison indicates that using Eq. (4) to calculate the value of $\alpha$ is more suitable to obtain the ${\rm H}\alpha$ lightcurves and time lags for the broadband PRM.

\begin{figure}[htb!]
\includegraphics[width=1.0\linewidth]{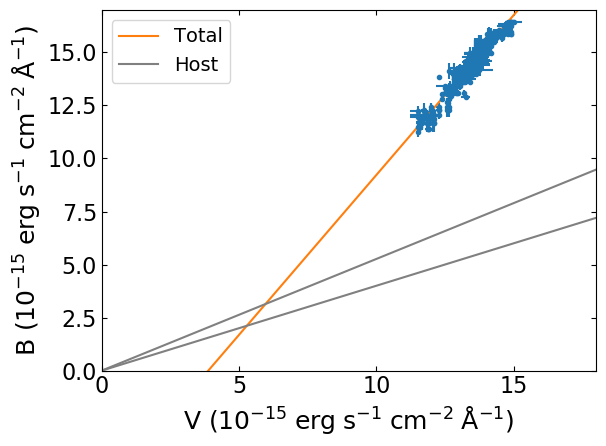}
\caption{The $B$ versus $V$ band fluxes of NGC 5548. A linear least-squares fit to data points yields the AGN slope plotted by the orange line. The range of host slopes plotted by two grey lines is taken from \citet{2010ApJ...711..461S}.}
\end{figure}

We also consider the influence of the host galaxy. The luminosity of the host galaxy changes little for the RM campaign durations, so it can be regarded as constant for most SRM campaigns. The shape of the lightcurves does not change with the host galaxy contribution for SRM and narrow band PRM\citep{2013A&A...552A...1P,2018A&A...620A.137R}. But for broadband PRM, the contribution of the host galaxy can change the value of $\alpha$ and then affect the shape of the extracted H$\alpha$  lightcurves. Its contribution must be considered.
We use the flux variation gradient \citep[FVG;][]{1981AcA....31..293C, 1992MNRAS.257..659W, 2012A&A...545A..84P} to determine the contribution of the host galaxy. As shown in Figure 3, we obtain the host galaxy contributions in the $B$ and $V$ bands for NGC 5548. They are $2.6\pm 0.5$ and $5.6\pm 0.3$ in units of $\rm 10^{-15}\;erg\;s^{-1}\;cm^{-2}\;\AA^{-1}$ for the $B$ and $V$ bands respectively, which are consistent with the result in \citet{2016ApJ...821...56F} ($2.88\pm 0.05$ and $5.25\pm 0.10$). 

The host galaxy contributions to the $g$, $r$ and $R$ bands can be estimated from the $B$ and $V$ band values by using the host galaxy template \citep{2007ApJ...663...81P}. A similar method is also applied to other 3 AGNs to estimate the host contributions to the continuum and line bands. 

\subsection{The JAVELIN Method}

\begin{figure}[htb!]
    \includegraphics[width=1.0\linewidth]{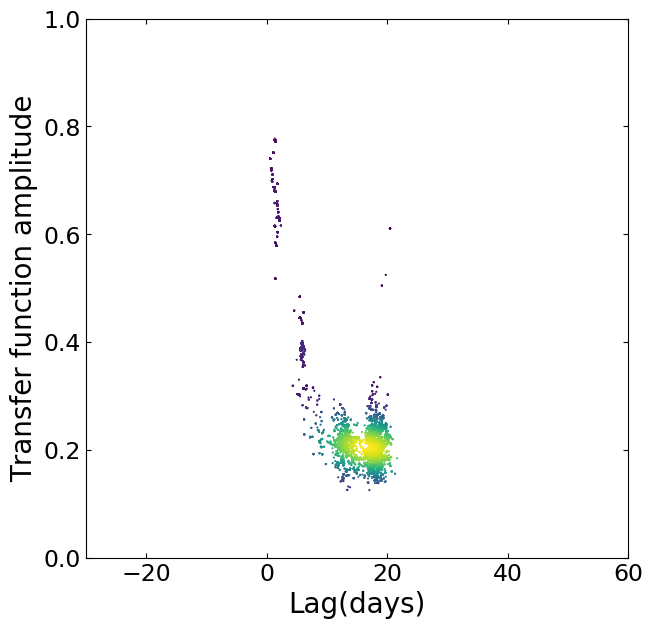}
    \caption{The time lags and the transfer function amplitude distribution of the JAVELIN Pmap model results for MCG +8-11-011 whose SRM ${\rm H}\beta$ lag is $15.72_{-0.52}^{+0.50}$ days. The color represents the number density of the points. The JAVELIN H$\alpha$ lag is $17.3_{-5.2}^{+1.5}$ days.}
\end{figure}

To check the reliability of the time lag calculated by the ICCF-Cut method, we also use another method to calculate the time lags of AGNs as the comparison. The Just Another Vehicle for Estimating Lags In Nuclei \citep[JAVELIN:][]{2011ApJ...735...80Z,2013ApJ...765..106Z,2016ApJ...819..122Z} program assumes that the lightcurve of AGNs can be modeled by a damped random walk (DRW) process and uses thousands of Markov Chain Monte Carlo (MCMC) DRW processes to get the distributions of the parameters including the time lag. Because the two-band photometry model (Pmap Model) of JAVELIN may be competitive with the SRM for strong (large equivalent width) lines such as ${\rm H}\alpha$ and ${\rm H}\beta$ \citep{2016ApJ...819..122Z}, we can use JAVELIN to calculate the time lag of AGNs with strong ${\rm H}\alpha$ emission lines. From the time lag distribution of JAVELIN, we use the highest posterior density to identify the time lag. The $1\sigma$ limits of the time lag that encompasses $68\%$ of the time lag distribution are adopted to obtain the upper and lower limits of the most probable time lag. To make sure the results of JAVELIN are reliable, in addition to the lag distribution, the transfer function amplitude between the continuum and the ${\rm H}\alpha$ emission line in the DRW model is also examined for the broadband PRM. The transfer function amplitude can represent the statistic mean of the ${\rm H}\alpha$ line contribution in the line band. The lag with much higher transfer function amplitude than the real ${\rm H}\alpha$ line ratio in the spectrum is not physically possible. Combining the distribution of lags and the transfer function amplitude, we can evaluate the reliability of the lag results. 

\begin{figure*}[!htb]
    \includegraphics[width=1.0\linewidth]{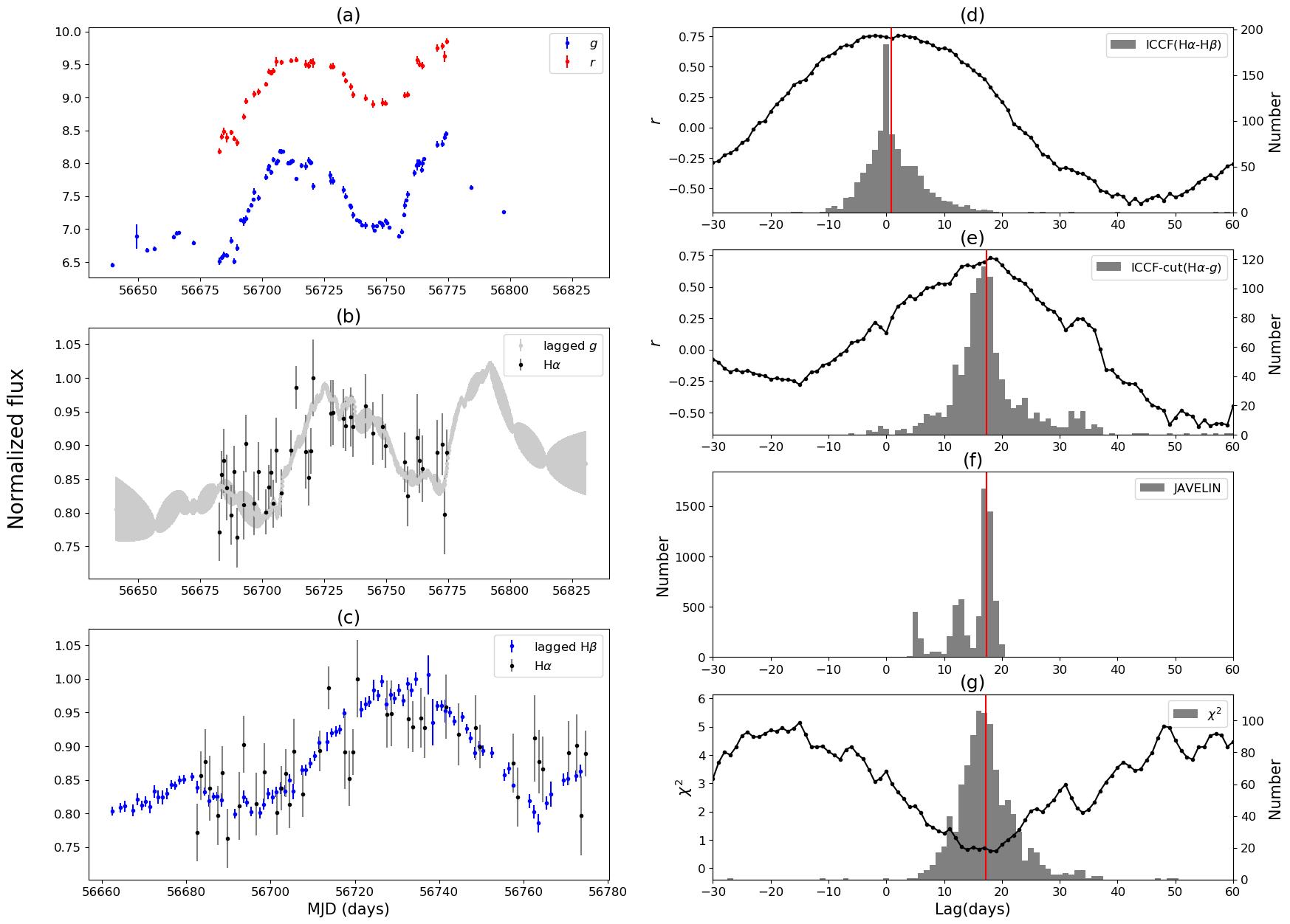}
    \caption{The lightcurves and lag distributions for MCG +8-11-011. The panel (a) shows the lightcurves of the continuum band ($g$) and line band ($r$). The panels (b) and (c) show the extracted H$\alpha$ lightcurves compared with the lagged continuum band and SRM H$\beta$ broad line lightcurves. The panel (d) shows the lag distribution between the SRM H$\beta$ line and extracted H$\alpha$ line. The panels (e), (f) and (g) show the lag distributions of the continuum band and extracted H$\alpha$ line with the ICCF-Cut, JAVELIN and $\chi^2$ methods respectively. The red line represents the median value of the lag distribution. For two ICCF results, the black lines represent the relations between the cross-correlation coefficient $r$ and the time lag. For the $\chi^2$ results, the black line represents the relation between the $\chi^2$ value and the time lag. The grey parts of these three panels (d),(e) and (g) represent the 1000 FR/RSS simulations. The grey part of JAVELIN (panel (f)) represents the distributions of 10000 MCMC simulations.}
\end{figure*}

Figure 4 shows the time lags and the transfer function amplitude distribution of the JAVELIN Pmap model results for MCG +8-11-011. The transfer function amplitude (on average $\sim 23\%$) can represent the ${\rm H}\alpha$ line ratio in the line band. It is consistent with the real ${\rm H}\alpha$ line ratio ($26\%$) calculated from the single-epoch spectrum. MCG +8-11-011 has the SRM ${\rm H}\beta$ lag of $15.72_{-0.52}^{+0.50}$ days, while the JAVELIN ${\rm H}\alpha$ time lag of $17.3_{-5.2}^{+1.5}$ days is consistent with the ICCF-cut ${\rm H}\alpha$ result of $17.3_{-4.3}^{+6.2}$ days. The ${\rm H}\alpha$ time lag is slightly larger than the SRM ${\rm H}\beta$ time lag, which is consistent with the structure of BLR and the results of previous works \citep{2000ApJ...533..631K,2010ApJ...716..993B,2012ApJ...755...60G}. We notice that the lags of a few points are close to zero with much higher transfer function amplitude (see Figure 4), which may be because the JAVELIN MCMC processes need higher line ratio as the time lag decreases to reproduce the given line band flux. To reduce such influence, we exclude the points with the transfer function amplitude larger than 0.4 and the lag results very close to zero.

\subsection{The $\chi^2$ Method}

\begin{figure*}[!htb]
    \includegraphics[width=1.0\linewidth]{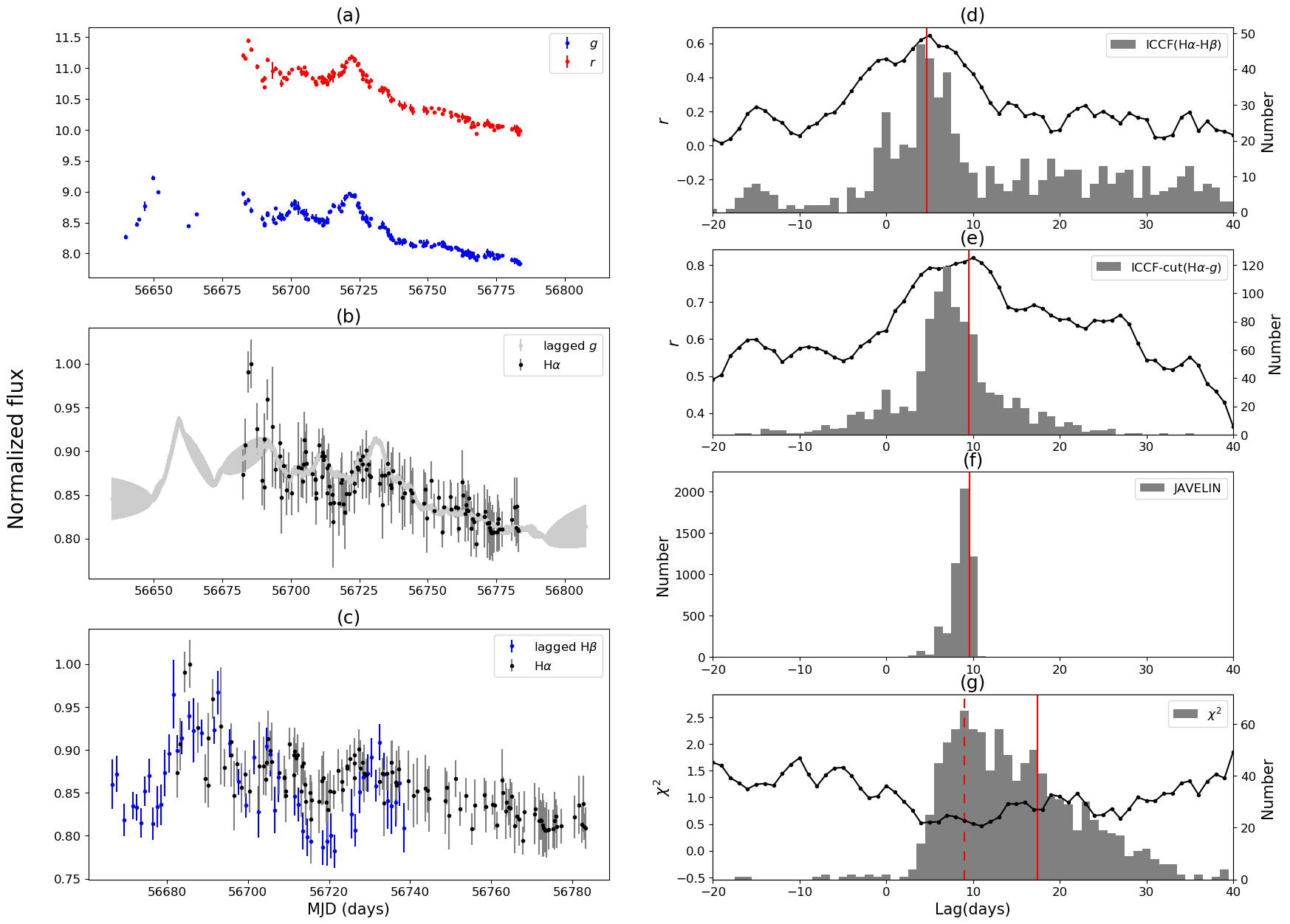}
    \caption{Same as Figure 6 but for NGC 2617. The red dash line in the panel (g) represents the peak value of the $\chi^2$ lag distribution. The red soild lines represent the median lags of the distributions.}
\end{figure*}

From the comparison of the lightcurves in panels (b) and (c) in Figure 5, it can be noticed that the errors of the extracted H$\alpha$ are much larger than the errors of the photometric broad bands and SRM H$\beta$ lightcurves. To evaluate the influence of errors and the feasibility of the broadband PRM with large errors, we also apply the $\chi ^2$ method \citep{2013A&A...556A..97C,2022ApJS..262...14B} which works better than using ICCF for AGNs with red-noise variability to obtain the H$\alpha$ time lag from the broadband and extracted H$\alpha$ lightcurves.

The $\chi ^2$ method uses the uncertainties to weight the data points in lightcurves. The $\chi ^2$ is calculated by

\begin{equation}
\chi^2 (\Delta t)  = \frac{1}{N}\sum_{i=1}^n\frac{(x_i - A_{\chi^2}y_{i,\Delta t})^2}{\delta x_i^2 + A_{\chi^2}^2\delta y_{i,\Delta t}^2},
\end{equation}
where $x_i$ and $y_{i,\Delta t}$ are the continuum band flux and the extracted H$\alpha$ flux with shifted lag $\Delta t$, $\delta x_i$ and $\delta y_{i,\Delta t}$ are their uncertainties. We interpolate and shift the extracted H$\alpha$ flux with the time $\Delta t$. For each line flux with the shifted lag $\Delta t$, we can obtain the value of $\chi^2 (\Delta t)$. It leads to a relation between the $\chi^2(\Delta t)$ value and time lag $\Delta t$. This process is similar to ICCF. $A_{\chi^2}$ is a normalized factor formulated as
\begin{equation}
    A_{\chi^2} = \frac{S_{xy} + (S_{xy}^2 + 4S_{x3y}S_{xy3})^{1/2}}{2S_{xy3}},
\end{equation}
where the coefficients are given by
\begin{equation}
\begin{aligned}
S_{xy}  &= \sum_{i=1}^{N}(x_i^2 \delta y_{i,\Delta t}^2 - y_{i,\Delta t}^2\delta x_i^2), \\
S_{xy3} &= \sum_{i=1}^{N}x_i y_{i,\Delta t} \delta y_{i,\Delta t}^2, \\
S_{x3y} &= \sum_{i=1}^{N}x_i y_{i,\Delta t}\delta x_i^2.
\end{aligned}
\end{equation}
We take the minimum points in the $\chi^2$ functions as the H$\alpha$ time lag
measurements. 

\begin{figure*}[!htb]
    \includegraphics[width=1.0\linewidth]{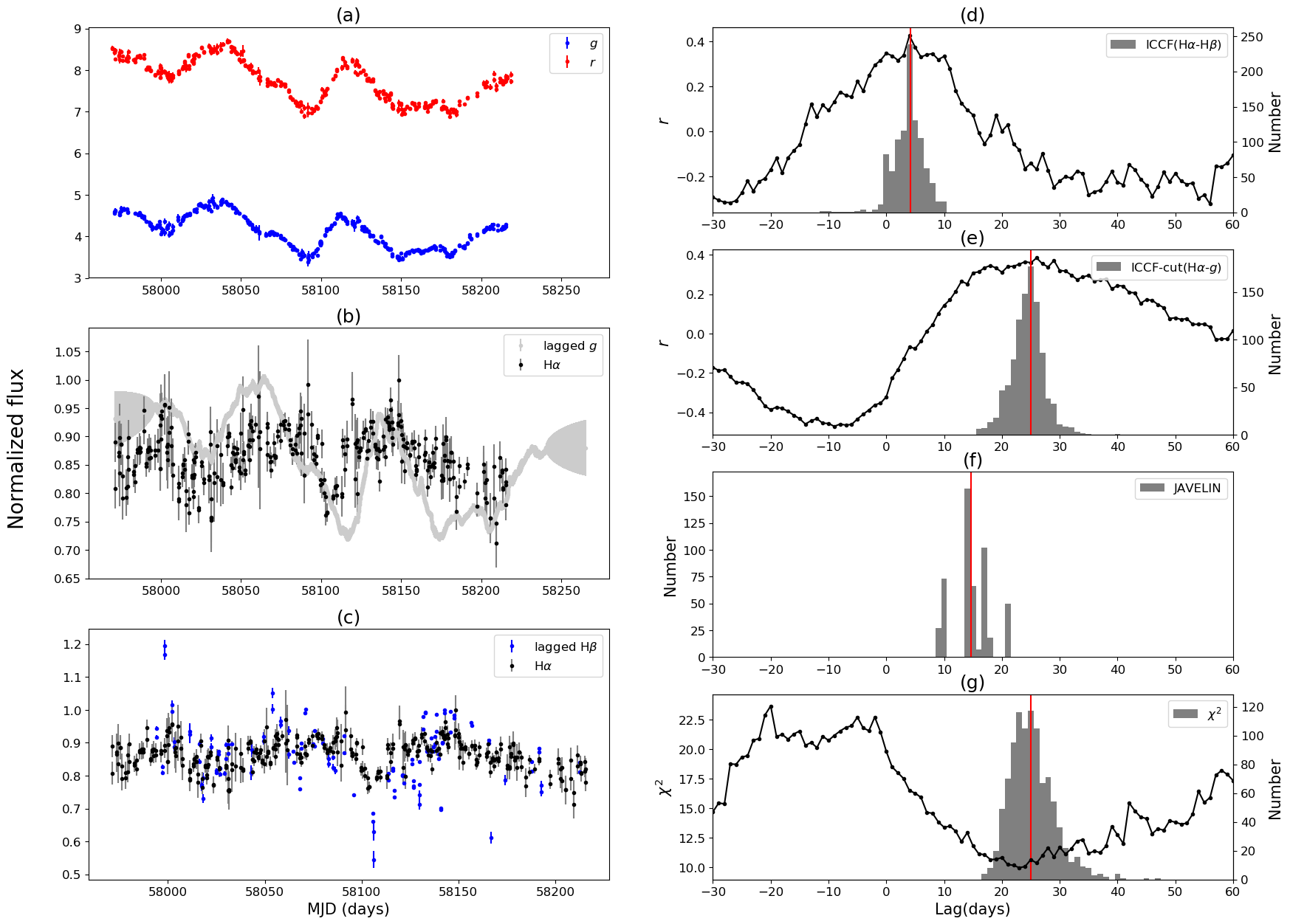}
    \caption{Same as Figure 6 but for 3C 120 without lightcurve segmentation.}
\end{figure*}

In order to estimate the uncertainties of the time lags of ICCF-Cut and the $\chi ^2$ method, we use the flux randomization (FR) and random subset selection (RSS) with Monte Carlo (MC) simulations \citep{1998ApJ...501...82P}. The FR alters the flux with the errors. Each data point is modified by adding a random noise according to a Gaussian distribution around the measured value, with the standard deviation of the measurement uncertainty. The RSS is used to estimate the errors of the unevenly sampled data by randomly excluding data points from the simulated lightcurves. Each realization is based on a randomly chosen subset of the original lightcurve. The FR procedure examines the sensitivity of the flux accuracy, and the RSS checks the effect of the incomplete sampling. For each target, we perform 1000 MC simulations to get the lag uncertainty of ICCF-Cut and the $\chi ^2$ method (see panel (g) in Figure 5). The $\chi ^2$ centroid ${\rm H}\alpha$ time lag of $17.2_{-3.9}^{+4.8}$ days for MCG +8-11-011 is consistent with the previous ICCF-Cut and JAVELIN results, which indicates that although the uncertainties of the extracted H$\alpha$ line flux are large compared with the variabilities of the lightcurves, the broadband PRM can still obtain the ${\rm H}\alpha$ time lag for these targets. The consistency of the H$\alpha$ lag distributions from the ICCF-Cut, JAVELIN and $\chi^2$ methods can ensure the reliability of the H$\alpha$ broadband PRM.

\section{Results for 4 Seyfert 1 Galaxies}

\begin{figure*}[!htb]
    \includegraphics[width=1.0\linewidth]{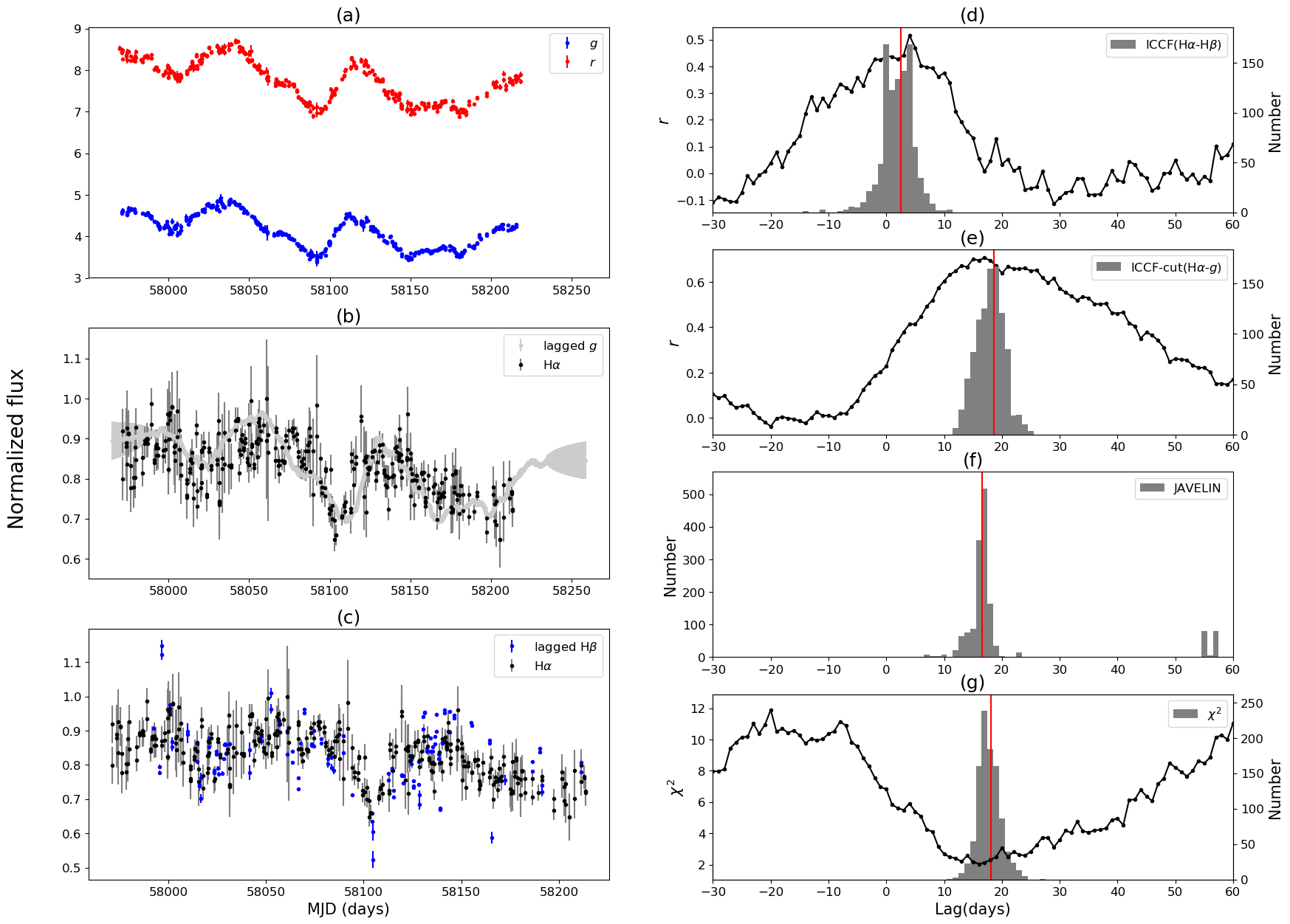}
    \caption{Same as Figure 6 but for 3C 120 with lightcurve segmentation into two parts at MJD 58092.}
\end{figure*}

We apply these methods to four Seyfert 1 galaxies. MCG +8-11-011 shows the best results, as shown in Figure 5. Besides the H$\alpha$ lag calculated in the previous part, we also use the ICCF to calculate the time lag between the SRM H$\beta$ and extracted H$\alpha$ light curve. (panel (d) in Figure 5). The high value of coefficient $r$ and a small lag between the SRM H$\beta$ and subtracted H$\alpha$ light curves confirm the reliability of the subtracted H$\alpha$ light curves.
Because the H$\alpha$ ratio obtained from the single-epoch spectrum is the highest among 4 galaxies and the lightcurves have obvious variabilities, the results of MCG +8-11-011 are better than other targets. The extracted H$\alpha$ lightcurve is well consistent with the lagged continuum and the lagged SRM H$\beta$ lightcurves. All the methods, including the ICCF-Cut, JAVELIN and $\chi^2$, show similar lag distributions for the continuum and extracted H$\alpha$ lightcurves. The H$\alpha$ lag is around 17 days.

\begin{figure*}[!htb]
    \includegraphics[width=1.0\linewidth]{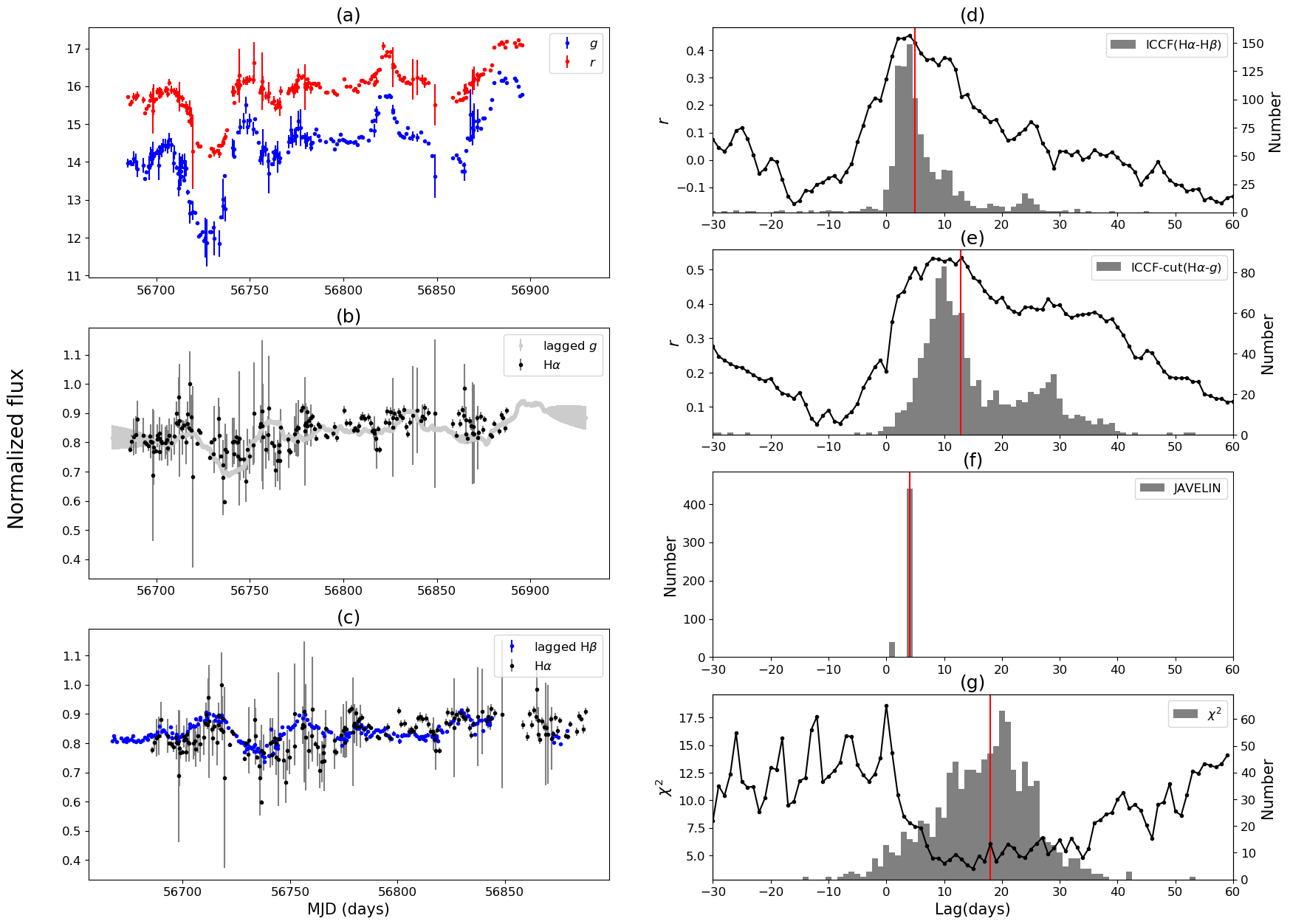}
    \caption{Same as Figure 6 but for NGC 5548 with the $g$ and $r$ band lightcurves with lightcurve segmentation into two parts at MJD 56772.}
\end{figure*}

For NGC 2617 shown in Figure 6, the extracted H$\alpha$ lightcurve is consistent with the lagged continuum and the lagged SRM H$\beta$ line lightcurves. The H$\alpha$ lag distributions of the ICCF-Cut and JAVELIN are very close. Because the variabilities are smaller than MCG +8-11-011 compared with uncertainties, the $\chi^2$ result is worse than those of ICCF-Cut and JAVELIN. Although the median lag of the $\chi^2$ method is much larger than the results of ICCF-Cut and JAVELIN, the peak value $10.0_{-0.2}^{+15.4}$ days of the $\chi^2$ method is very close to the results of ICCF-Cut and JAVELIN. Because NGC 2617 was discovered by \citet{2014ApJ...788...48S} to be a changing look AGN, we abandon the SRM H$\beta$ line lightcurve with MJD$>56735$  \citep{2017ApJ...840...97F} for the comparisons. The latter part of the lightcurves with lower fluxes may also be one of the reasons for the bad $\chi^2$ result.

\begin{figure*}[!htb]
    \includegraphics[width=1.0\linewidth]{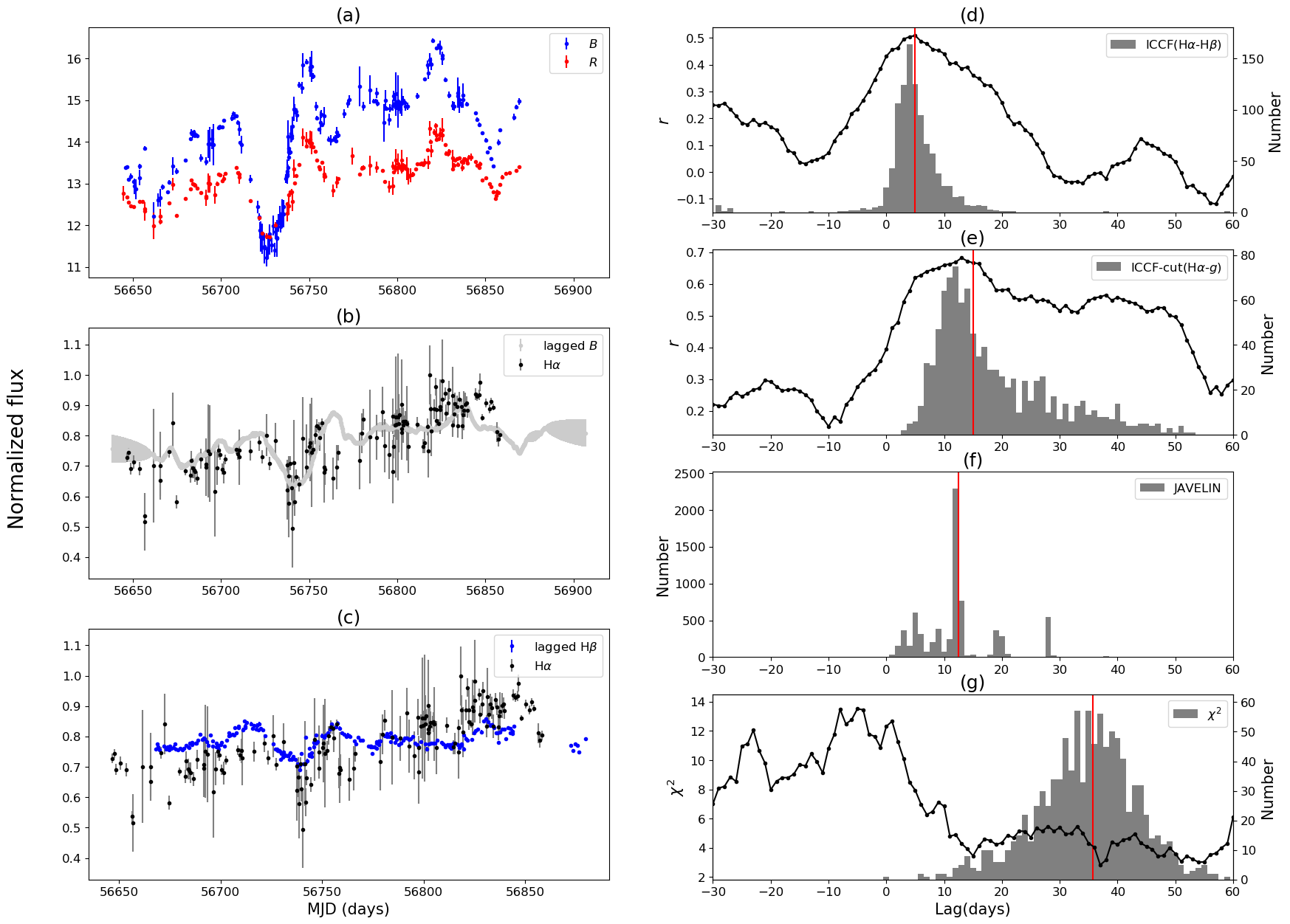}
    \caption{Same as Figure 6 but for NGC 5548 with the $B$ and $R$ band lightcurves with lightcurve segmentation into two parts at MJD 56766.}
\end{figure*}

We apply the same methods to 3C 120, as shown in Figure 7. But the extracted H$\alpha$ line lightcurve is not well consistent with the continuum band (panel (b) in Figure 7). The left part of the extracted H$\alpha$ line lightcurve is obviously lower than the continuum band lightcurve and the right part is higher. The correlation coefficient value of ICCF-Cut is also very low. We noticed that even for the simultaneous H$\beta$ SRM, the correlation coefficient value of ICCF between the continuum and H$\beta$ line is also not high, only about 0.4 \citep{2020MNRAS.497.2910H}. Another issue is that the observational duration of 3C 120 is about 250 days, twice of the durations of MCG +8-11-011 and NGC 2617. The spectral index of the AGN continuum usually changes little within several months as in the cases of MCG +8-11-011 and NGC 2617, but for 3C 120, the spectral index of the continuum and the value of $\alpha$ may change during the longer observation durations. To understand the deviation between the extracted H$\alpha$ line and the continuum, we adjust Equation (4) to calculate the value of $\alpha$. We divide the lightcurve into two parts so that each part of the lightcurve has similar duration time as MCG +8-11-011 and NGC 2617, and for each part the value of $\alpha$ is adjusted according to its average line band and continuum band fluxes. For each part of the lightcurves, the $\alpha_i$ is calculated by
\begin{equation}
\alpha_i=\alpha \frac{\overline{L_{\rm cont}}}{\overline{L_{\rm line}}} \frac{\overline{L_{{\rm line},i}}}{\overline{L_{{\rm cont},i}}},
\end{equation}
where $\overline{L_{\rm cont}}$ and $\overline{L_{\rm line}}$ are the average fluxes in the the continuum and line bands for the whole period lightcurves, while $\overline{L_{{\rm cont},i}}$ and $\overline{L_{{\rm line},i}}$ are the average fluxes for each part of lightcurves. 

Figure 8 shows that for 3C 120 the subtracted ${\rm H}\alpha$ lightcurve with the varying $\alpha$ value in two duration parts (separated at MJD 58092) is more consistent with the lagged continuum band lightcurve. Similar to the ICCF-Cut, we also divide the initial lightcurves into two parts to calculate the JAVELIN lag respectively, then combine the two part results into the final one. All three methods show similar lag distributions in Figure 8. Although the ICCF-Cut ${\rm H}\alpha$ lag of $18.6_{-2.7}^{+2.2}$ days is slightly shorter than the SRM H$\beta$ lag of $21.2_{-1.0}^{+1.6}$ days, the time lag between the SRM H$\beta$ line and the extracted ${\rm H}\alpha$ line calculated by ICCF still shows that the extracted H$\alpha$ lightcurve is possibly lagged behind the SRM H$\beta$ lightcurve with $2.5_{-3.0}^{+2.3}$ days (panel (d) in Figure 8). This contradiction may be due to the lower accuracy of the SRM H$\beta$ lightcurve compared with other targets. This contradiction may also be ignored because the lag uncertainties are larger than the difference between the H$\alpha$ and H$\beta$ lags. The extracted ${\rm H}\alpha$ lag is consistent with the SRM H$\beta$ lag in general, which has also been found by the previous SRM research of 3C 120 showing that its ${\rm H}\alpha$ lag ($28.5_{+9.0}^{-8.5} $ days) and H$\beta$ lag ($27.9_{+7.1}^{-5.9} $ days) are very close \citep{2014A&A...566A.106K}. The consistency of H$\alpha$ lightcurve with the lagged continuum and SRM H$\beta$ lightcurves as well as the similar lag distributions for three methods indicate that this $\alpha$ value adjustment method is effective for the broadband PRM with longer duration time.

For NGC 5548 with more than 200 days observational duration, we also divide the lightcurves into two parts (separated at MJD 56772 in Figure 9). The extracted H$\alpha$ lightcurve is consistent with the lagged continuum band and the lagged SRM H$\beta$ line lightcurves in general. Because the flux uncertainties of the continuum and line bands are larger than other targets, the lag distributions of three methods are not very consistent with each other. The lag value of JAVELIN is much smaller than others. We will use the simulations to explain such a difference in Section 5.
To determine the H$\alpha$ lag and examine the reliability of the broadband PRM, besides the $g$ and $r$ bands, we also used the $B$ band as the continuum band and the $R$ band as the line band (separated at MJD 56766 in Figure 10). The results of the $B$ and $R$ bands are similar to the results of the $g$ and $r$ bands, especially for the lag distributions of ICCF-Cut. Although the result of the $\chi^2$ method is worse, the lag distributions of the ICCF-Cut and JAVELIN are still similar. 

Considering the simultaneous SRM H$\beta$ lag of $4.17_{-0.36}^{+0.36}$ days \citep{2017ApJ...837..131P} as the broadband PRM and the SRM H$\alpha$ lag of $11.02_{-1.15}^{+1.27}$ days in other period \citep{2010ApJ...716..993B} for NGC 5548, the lag distributions in Figure 9 and Figure 10 are probably reasonable. Especially the lag distributions of the ICCF-Cut and $\chi^2$ method for the $g$ and $r$ bands and the lag distributions of the ICCF-Cut and JAVELIN for the $B$ and $R$ bands are more reliable and consistent with each other.

\begin{figure*}
\includegraphics[width=1.0\linewidth]{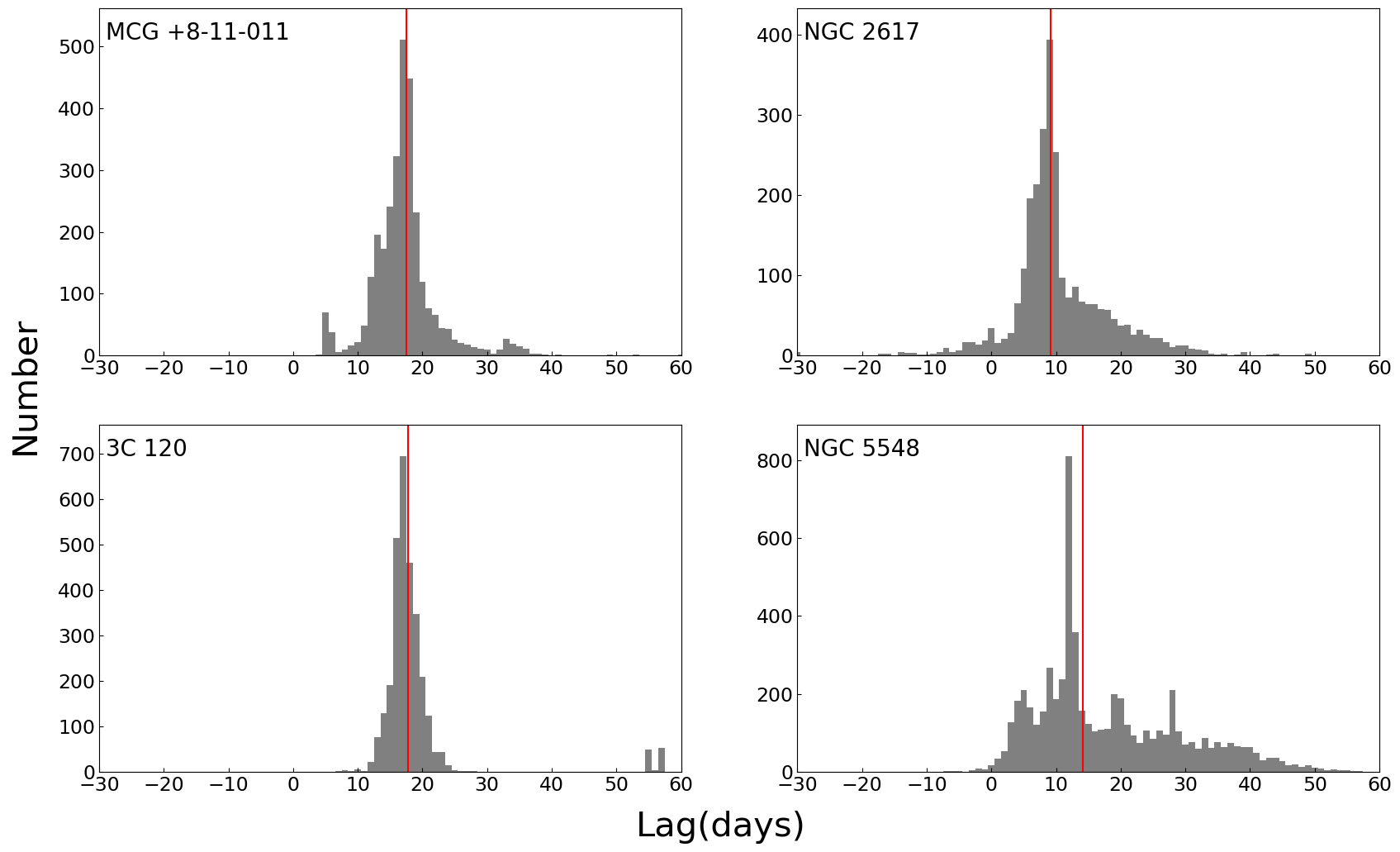}
\caption{The combined lag distributions for three methods with the same weight. The red lines represent the median lags. For NGC 5548, it contains the lag distributions of both $g,r$ and $B, R$ bands.}
\end{figure*}

\begin{deluxetable*}{ccccccc}
    \centering
    \tablecaption{The ${\rm H}\alpha$ lag results (in days) of 4 Seyfert 1 galaxies. \label{tab:mathmode}}
    \tablehead{
    \colhead{Name}  & \colhead{ICCF-Cut} & \colhead{JAVELIN} & \colhead{$\chi^2$} & \colhead{Combined} & \colhead{SRM ${\rm H}\beta$} &  \colhead{H$\beta$ vs H$\alpha$}
    }
    \startdata
    MCG +8-11-011 & $17.3_{-4.3}^{+6.2}$ & $17.3_{-5.2}^{+1.5}$ & $17.2_{-3.9}^{+4.8}$ &  $17.4_{-3.9}^{+4.8}$ & $15.72_{-0.52}^{+0.50}$ & $0.8_{-4.1}^{+5.4}$ \\
    NGC 2617 & $9.5_{-7.0}^{+9.1}$ & $9.6_{-0.6}^{+0.7}$ & $17.5_{-7.7}^{+7.9}$ & $9.0_{-12.3}^{+7.4}$ & $4.32_{-1.35}^{+1.10}$ & $4.7_{-6.3}^{+3.3}$ \\
    3C 120 & $18.6_{-2.7}^{+2.2}$ & $16.6_{-3.9}^{+2.6}$ & $18.1_{-1.5}^{+2.0}$ & $17.8_{-1.6}^{+2.5}$ & $21.2_{-1.0}^{+1.6}$ & $2.5_{-3.0}^{+2.3}$ \\
    NGC 5548 ($gr$) & $12.8_{-4.8}^{+14.7}$ & $4.5_{-0.2}^{+0.1}$ & $18.0_{-9.3}^{+7.0}$ & 
    \multirow{2}{*}{$14.3_{-5.8}^{+16.9}$}  & $4.17_{-0.36}^{+0.36}$ & $5.0_{-2.5}^{+6.5}$ \\
    NGC 5548 ($BR$) & $15.0_{-5.1}^{+14.2}$ & $12.5_{-6.7}^{+7.2}$ & $35.7_{-8.8}^{+7.8}$ &  & - & $5.0_{-2.5}^{+4.0}$\\
    \enddata
\tablecomments{The H$\beta$ vs H$\alpha$ lags are obtained from the SRM H$\beta$ and the extracted H$\alpha$ lightcurves with ICCF. For NGC 5548, the combined lag is obtained from the lag distributions of both $g,r$ and $B, R$ bands.}
\end{deluxetable*}

\begin{figure}
\includegraphics[width=1.0\linewidth]{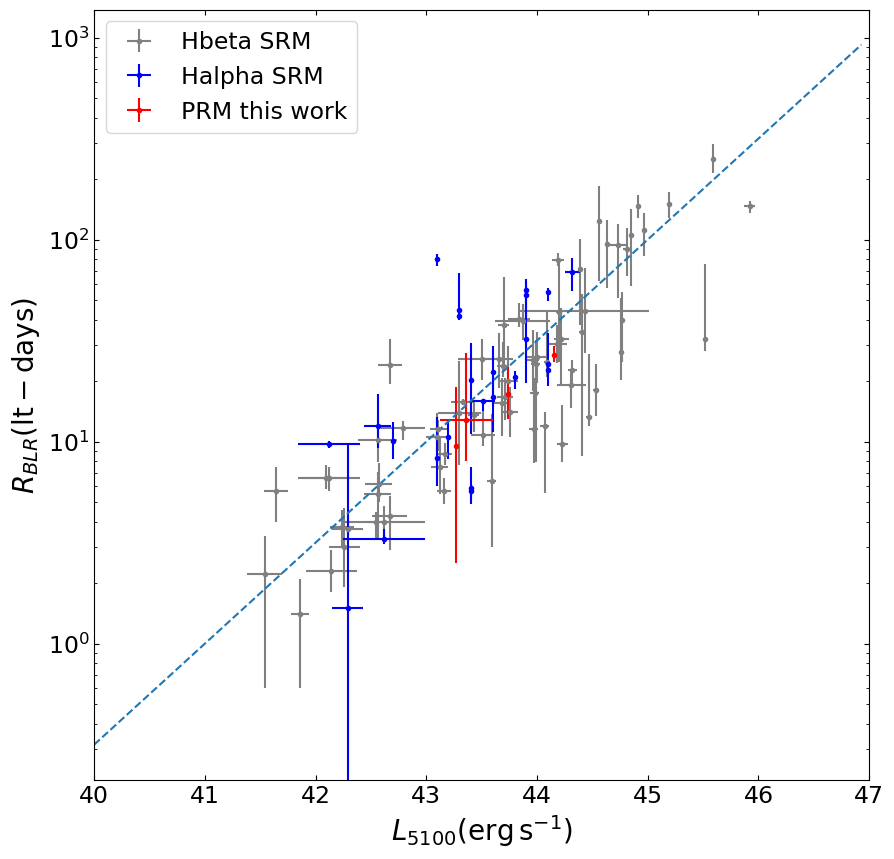}
\caption{The $R_{BLR}-L_{5100}$ relationship of the broadband PRM, SRM for ${\rm H}\alpha$ line (blue points) and SRM for ${\rm H}\beta$ line (black points) \citep{2019ApJ...886...42D}. The red points represent the ${\rm H}\alpha$ time lags obtained with the ICCF-Cut for the $g$ and $r$ bands. The dash line is the $R-L$ relation given by \citet{2019FrASS...6...75P}.}
\end{figure}
    
\begin{figure}
\includegraphics[width=1.0\linewidth]{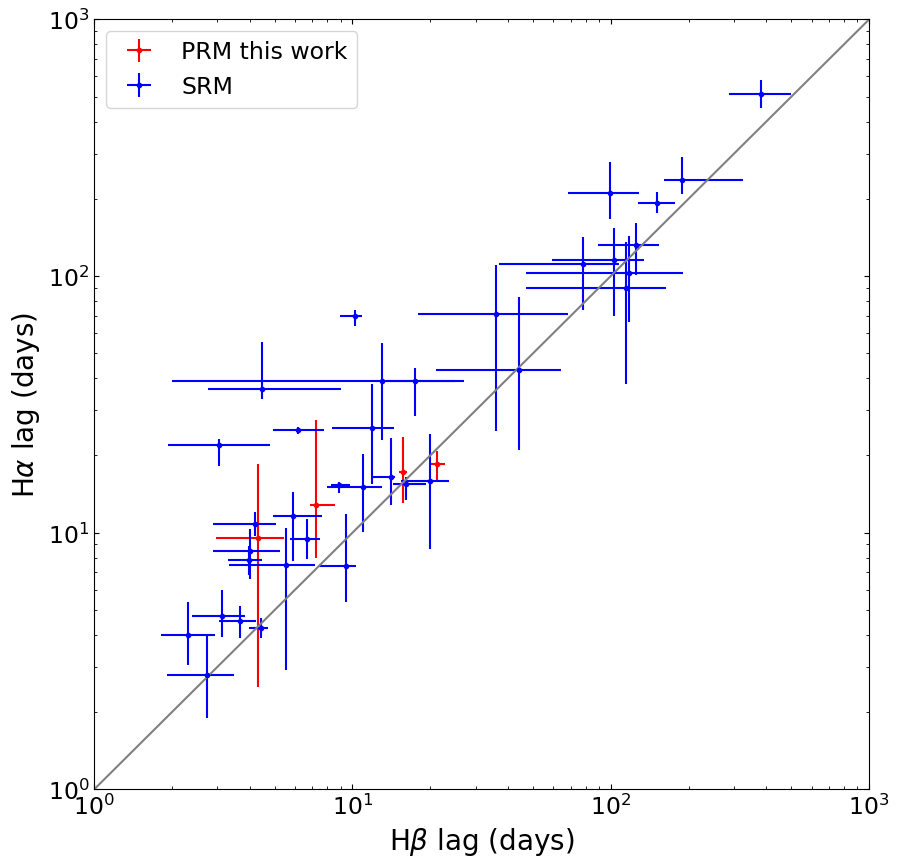}
\caption{A comparison of the ${\rm H}\alpha$ and ${\rm H}\beta$ time lags for 8 AGNs with SRM results. The red points are from this work and the blue points are from SRM \citep{2000ApJ...533..631K,2010ApJ...716..993B,2017ApJ...849..146G}. The solid line represents the one-to-one ratio.}
\end{figure}

All H$\alpha$ lag results (median values) for 4 Seyfert 1 galaxies are listed in Table 2. 
We try to plot the lag distribution of three methods with the same weight in one figure (Figure 11) and use the highest posterior density to obtain the lags as the comparison. We find that these combined lags are similar to the ICCF-Cut results. Because the cross-correlation function \citep[CCF;][]{1982ApJ...255..419B} method has been widely used and examined for decades in many RM projects, we chose the results of ICCF-cut as the final lags (using the combined lag results has no significant changes). It is also more convenient to use these lags to compare with other SRM results which are mainly obtained from the CCF and its variants.
We compare our broadband PRM ${\rm H}\alpha$ lags with the SRM time lags in the $R-L$ relation (see Figure 12). Our results from the ${\rm H}\alpha$ PRM are consistent with the commonly adopted $R_{\rm{BLR}} \varpropto L^{\alpha}$ relationship \citep{2019FrASS...6...75P}. We also compare our ${\rm H}\alpha$ lag results with those of SRM H$\beta$ lags. Figure 13 shows that on average the ${\rm H}\alpha$ time lags are slightly larger than the SRM H$\beta$ time lags, which is consistent with the standard model of AGNs, where the BLR size of ${\rm H}\alpha$ line is usually larger than that of H$\beta$ line \citep{2000ApJ...533..631K,2010ApJ...716..993B,2012ApJ...755...60G,2017ApJ...849..146G}.

\section{Discussion}

\begin{figure*}[htb!]
    \includegraphics[width=1.0\linewidth]{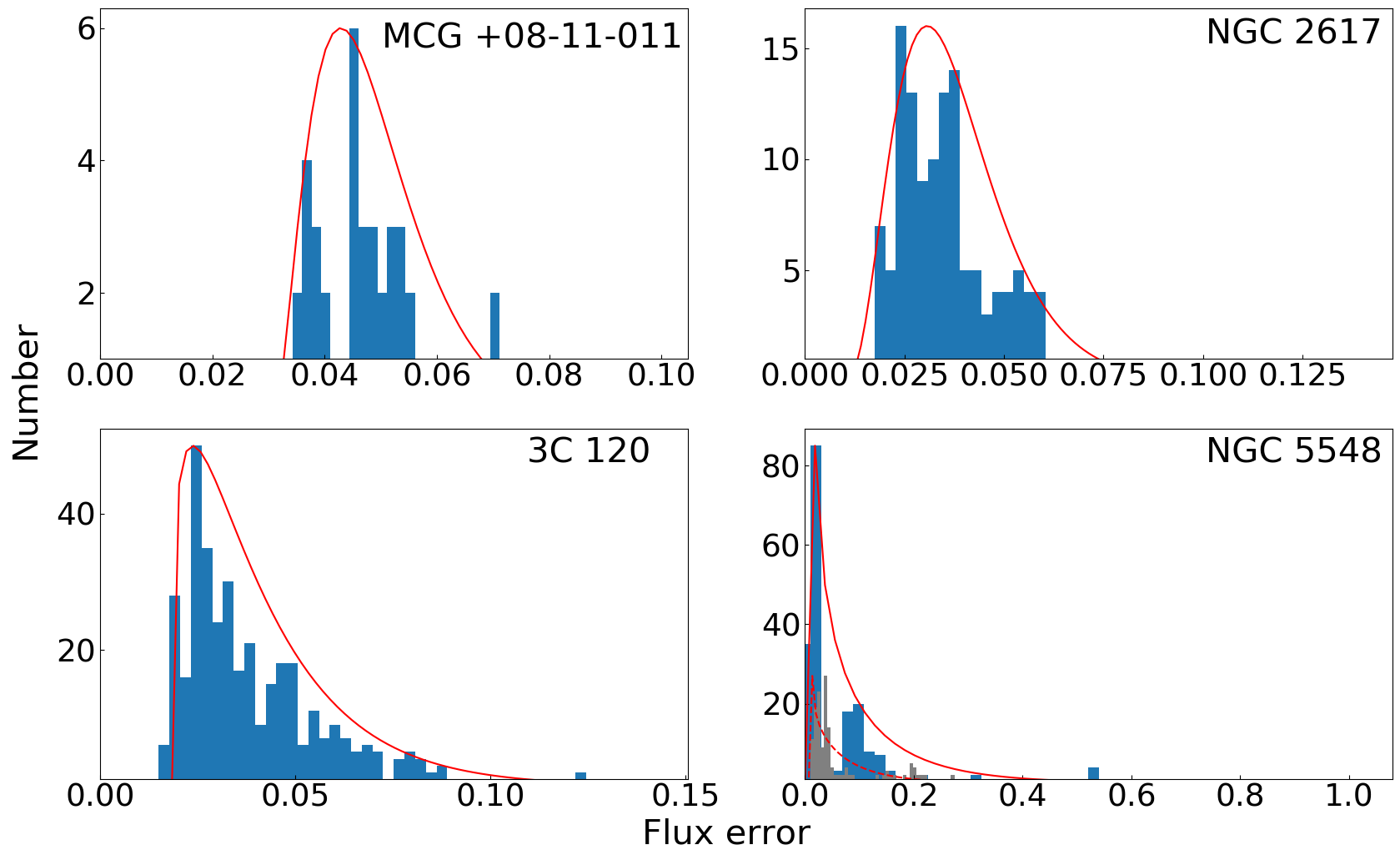}
    \caption{The flux error distributions of the line band for 4 Seyferts. The blue histograms represent the error distributions of the $r$ band and the red lines represent their fitting skewed normal distribution. For NGC 5548, the grey histogram and red dash line represent the data of $R$ band.}
\end{figure*}

\begin{figure*}[htb!]
\includegraphics[width=1.0\linewidth]{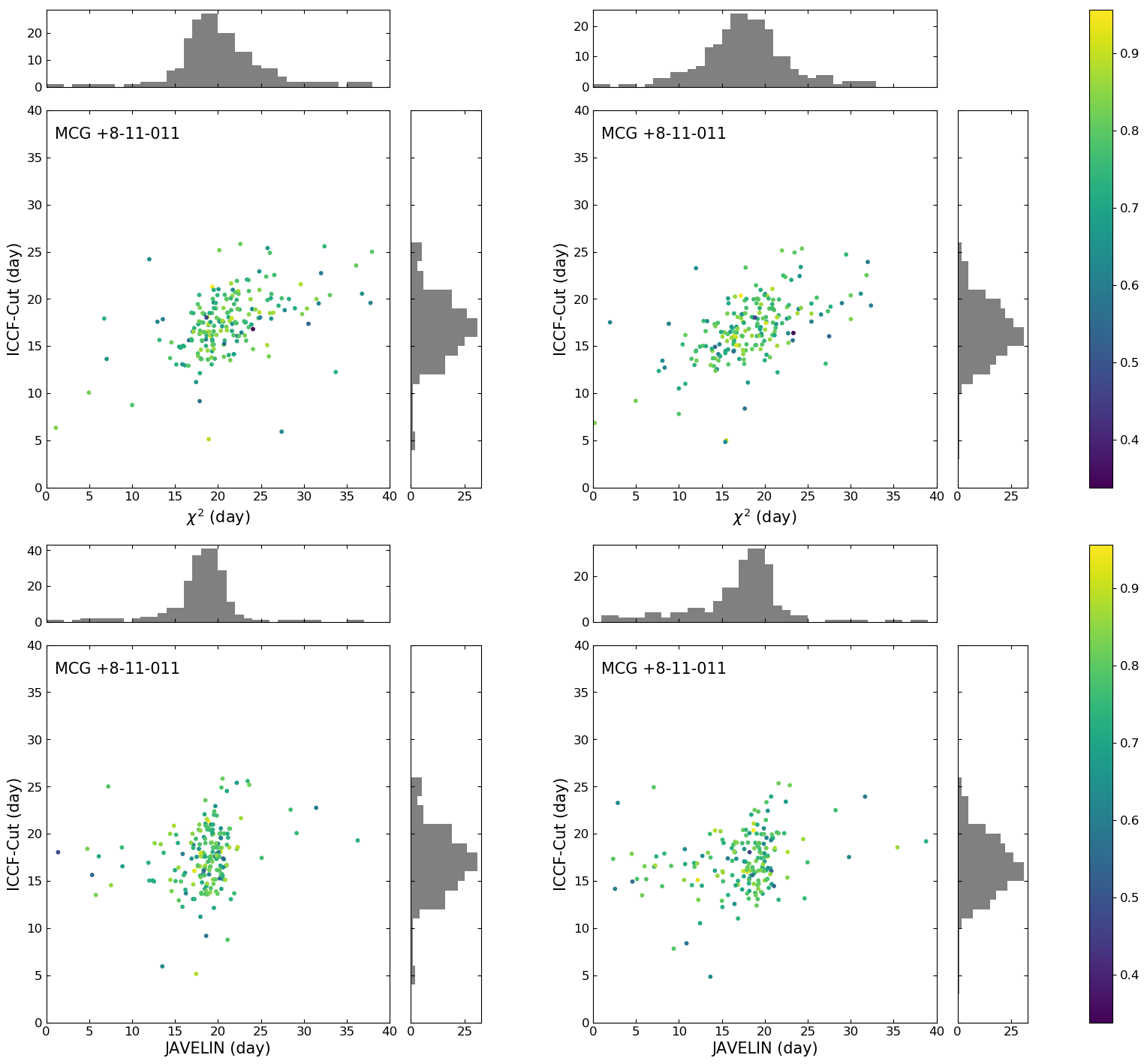}
\caption{The results of three methods for 200 mock lightcurves with the initial ${\rm H}\alpha$ time lag set as 20 days for MCG +8-11-011. The left panels show the simulations without ${\rm H}\beta$ line contribution in the continuum band, and the right panels show the simulations with ${\rm H}\beta$ line contribution in the continuum band and 1 day continuum lag in the line band. The colors of points represent the peak values of the cross-correlation coefficient $r$ for ICCF-Cut.}
\end{figure*}

To examine the reliability of the time lags and the influence of ${\rm H}\beta$ emission lines in the continuum band, we use the DRW model to produce the mock lightcurves of AGNs. The DRW process can be described by a stochastic differential equation \citep{2009ApJ...698..895K},

\begin{equation}
dc(t) = -\frac{1}{\tau}c(t)dt+{\sigma}\sqrt{dt}{\epsilon}(t)+b\tau,
\end{equation}
where $c(t)$ is the continuum flux, $\tau$ is the relaxation time of the continuum, $\sigma$ is the standard deviation of the continuum, and $\epsilon(t)$ is a white Gaussian noise process with zero mean and the variance equal to one. The mean value of the continuum is $b\tau$ and the variance is $\sigma{\tau}^2/2$. The variability of the broad emission line relative to the continuum can be described as

\begin{equation}
l(t)=\int {\Psi}(t-t')c(t)dt,
\end{equation}
where $\Psi (t)$ is the transfer function between the continuum and the broad emission line. We use a top hat (rectangular function) for the transfer function centered at a time lag $\tau_d$ with a width $w$ and an amplitude $A$,

\begin{equation}
\Psi(t)=\frac{A}{w} \quad {\rm for} \quad  \tau_d - \frac{w}{2} \leqslant t \leqslant \tau_d + \frac{w}{2}.
\end{equation}

\noindent Combining the transmission functions of the continuum and line bands and the ${\rm H}\alpha$ and ${\rm H}\beta$ line strengths, as well as the parameters of the AGN lightcurve variability, we can produce the mock lightcurves of the continuum and line bands.

We use JAVELIN to obtain the DRW parameters from the observational data of the four Seyfert 1 galaxies, and use these parameters to reproduce the mock lightcurves. We set the H$\alpha$ and H$\beta$ line strengths obtained from the spectra as the transfer function amplitude. To simulate the observational errors, we use the skewed normal distribution to fit the error distributions for 4 sources (see Figure 14) and use the same skewed normal distribution errors to reproduce the mock lightcurves. To simulate the small inter-continuum time lag between the continuum emissions in the continuum and line bands, the mock line band consists of the 1 day lagged continuum and 20 day lagged H$\alpha$ line.

\begin{figure*}[htb!]
    \includegraphics[width=1.0\linewidth]{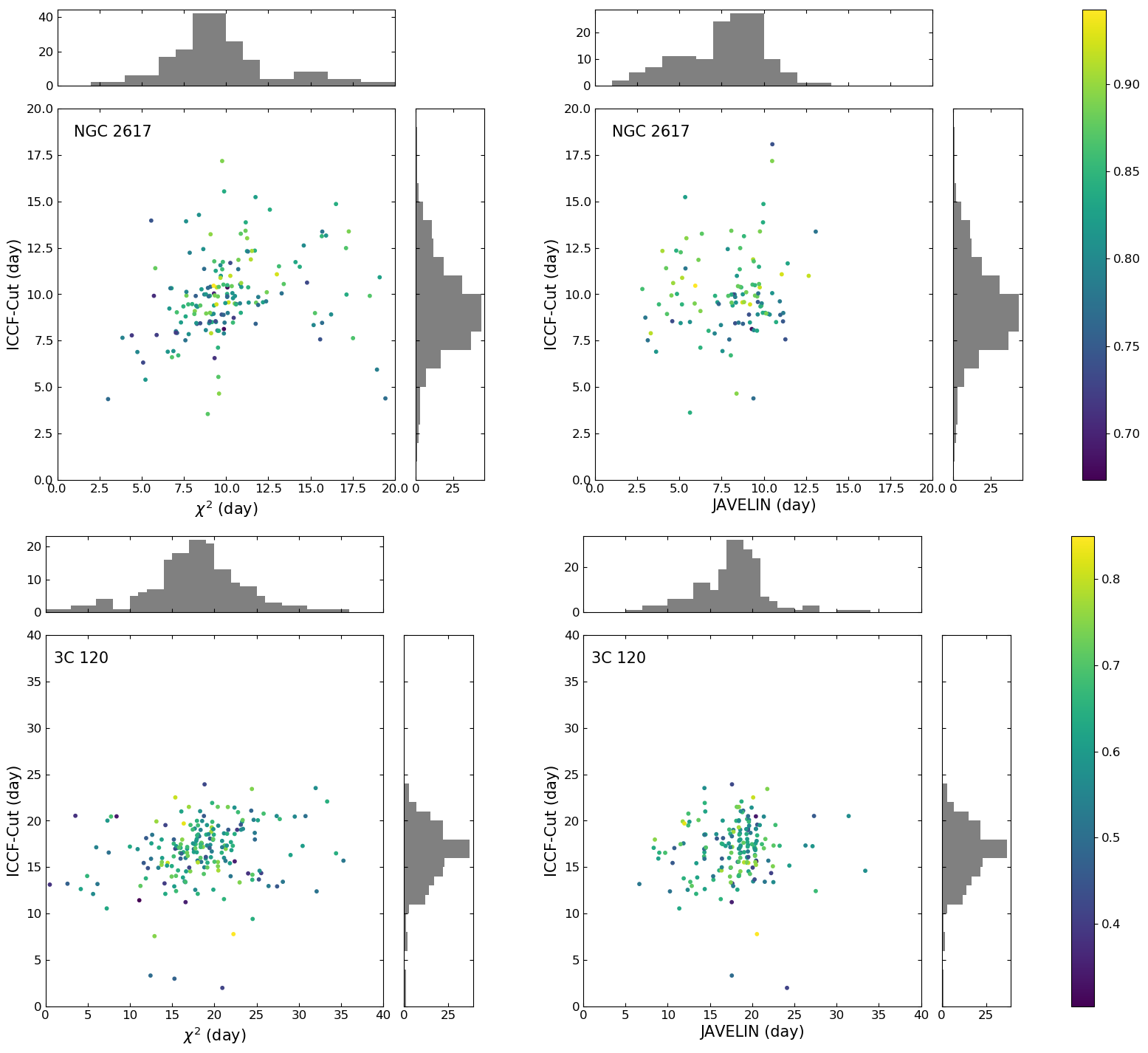}
    \caption{The results of three methods for NGC 2617 and 3C 120. The left panels show the results of the ICCF-Cut and $\chi^2$ methods, and the right panels show the results of the ICCF-Cut and JAVELIN. The colors of points represent the peak value of the cross-correlation coefficient $r$ for ICCF-Cut.}
\end{figure*}

To make sure the variability of mock lightcurves is similar to that of the sources, we use the Welch-Stetson J Variability Index \citep{1993AJ....105.1813W} to evaluate the variability. The J index is composed of the relative error ($\delta$), the normalized residuals of a pair of observations ($P_k$), and a weighting factor ($w_k$). The relative error is defined by \citet{1996PASP..108..851S} as
\begin{equation}
\delta_i= \frac{f_i-\overline{f}}{\sigma_{f,i}}\sqrt{\frac{n}{n-1} }.
\end{equation}
Here $n$ is the number of observations, $\sigma_{f,i}$ is the measurement error and $\overline{f}$ is the mean flux of the light curve. To reduce the influence of very large flux change within few data points, the weight factor is defined as
\begin{equation}
  w_i=\left[1+(\frac{\delta_i}{2})^2\right]^{-1}.
\end{equation}
The J index is defined as
\begin{equation}
  \rm J=\frac{\sum w_k sgn(\delta_i^2-1) \sqrt{\left\lvert \delta_i^2-1 \right\rvert } }{\sum w_k }.
\end{equation}
Here sgn simply returns the sign of the value. $\rm J<0$ means that the variability is dominated by the uncertainties of the observation. After reproducing the mock lightcurves with the DRW model, we select the mock lightcurves which have similar J index with the real observational data.

\begin{figure*}[htb!]
    \includegraphics[width=1.0\linewidth]{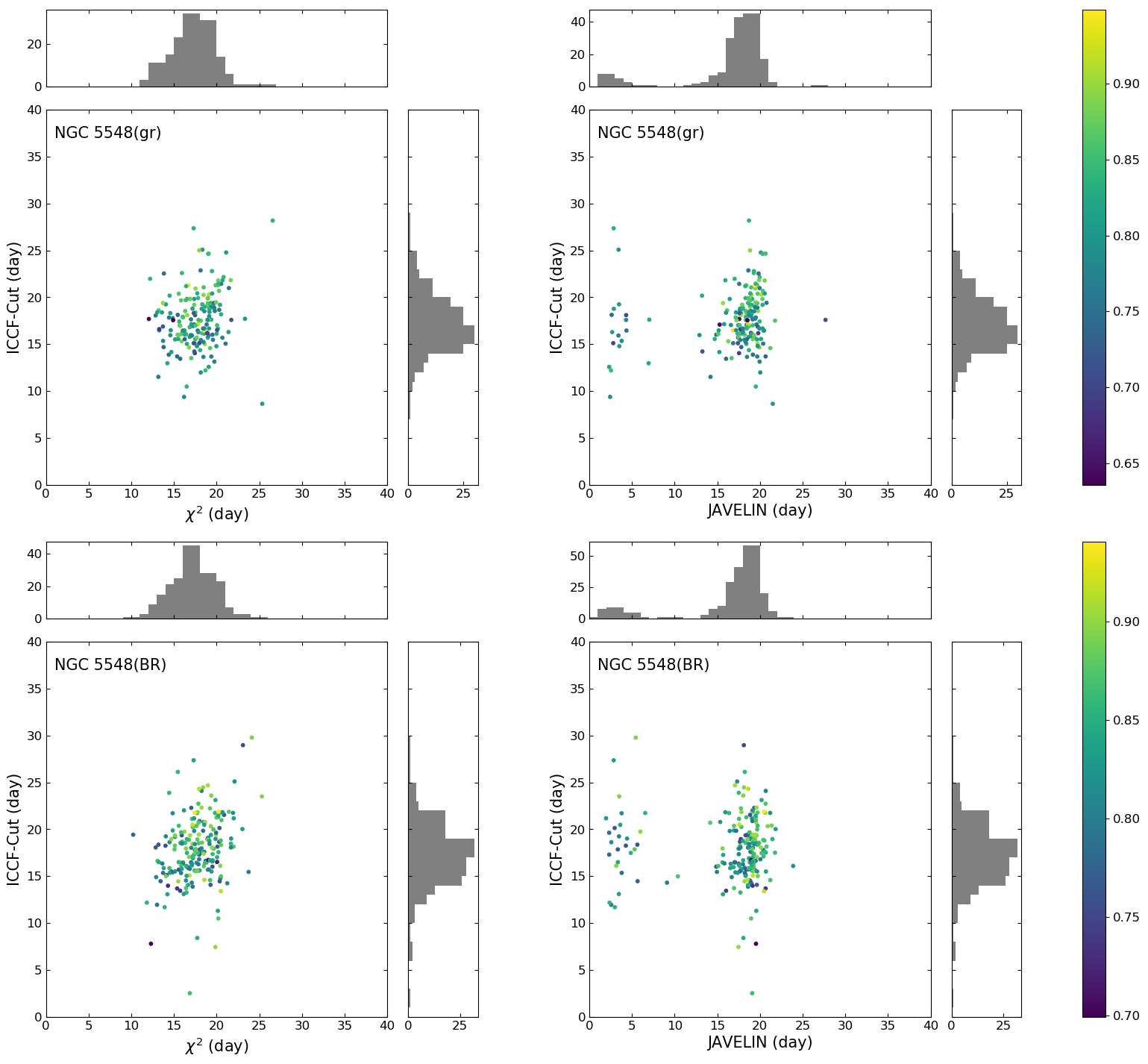}
    \caption{Same as Figure 16 but for NGC 5548. The two top panels represent the results of the $g$ and $r$ bands. The two bottom panels represent the results of the $B$ and $R$ bands.}
\end{figure*}

For each set of parameters, we use the DRW model to reproduce four mock lightcurves in one simulation. One pair of lightcurves are those of the (pure) continuum band continuum and the line band continuum with the ${\rm H}\alpha$ emission line. This pair of lightcurves represent the ideal data to calculate the time lag with the ICCF-Cut, JAVELIN and $\chi ^2$ methods. 
To simulate the small inter-continuum time lag between the continuum emissions in the continuum and line bands and evaluate the influence of the H$\beta$ line in the continuum band, the mock continuum band consists of the continuum and the 15 day lagged H$\beta$ line, the mock line band consists of the 1 day lagged continuum and 20 day lagged H$\alpha$ line for MCG +8-11-011.

\begin{figure*}[htb!]
    \plotone{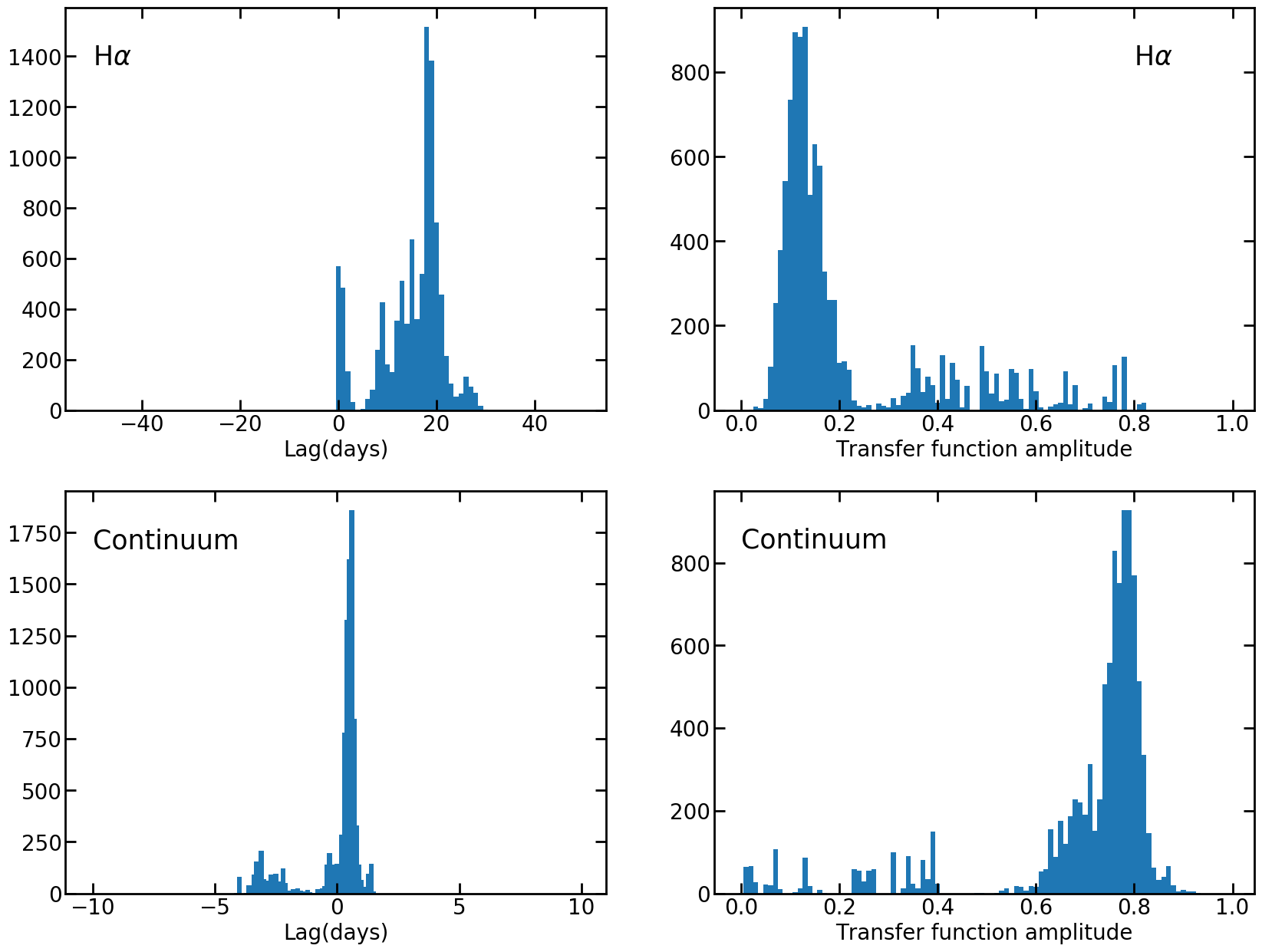}
    \caption{The JAVELIN DPmap model results for MCG +08-11-011. Left panels show the lag distributions and right panels show the line or continuum strength ratio distributions. The upper left panel represents the ${\rm H}\alpha$ lag and the bottom left panel represents the inter-continuum lag between the $g$ and $r$ bands. The initial lag limits are set to $-50\sim 50$ days for the broad emission line and $-10\sim 10$ days for the inter-continuum lag. The DPmap model lag is $18.4_{-6.0}^{+2.4}$ days and the inter-continuum lag is $0.5_{-0.3}^{+0.2}$ days.}
\end{figure*}

We simulate 200 pairs of the mock lightcurves which have similar parameters with MCG +8-11-011 such as the cadence, variabilities, ${\rm H}\alpha$ line strength in the line band, and ${\rm H}\beta$ line strength in the continuum band. Then we use three methods to calculate the time lags for each pair of the mock lightcurves and get the distributions of ${\rm H}\alpha$ time lag (see Figure 15). By comparing the left and right panels of Figure 15, we find that although the ${\rm H}\beta$ line and inter-continuum lag can slightly influence the lag distributions, most of the lags estimated by three methods are clustered around the true lag of 20 days, indicating that three methods are efficient for the broadband PRM and the influence of the ${\rm H}\beta$ emission line and the inter-continuum lag can be ignored for the H$\alpha$ lag calculations. From the top panels of Figure 15, it can be noticed that the results of the ICCF-Cut and $\chi ^2$ methods have positive correlation, which means that we may obtain the similar but not independent lag distributions with the ICCF-Cut and $\chi ^2$ methods. Only relying on the consistency of the results with the ICCF-Cut and $\chi ^2$ methods we may obtain biased results. We still need other methods to confirm the results.

\begin{figure*}[htb!]
    \plotone{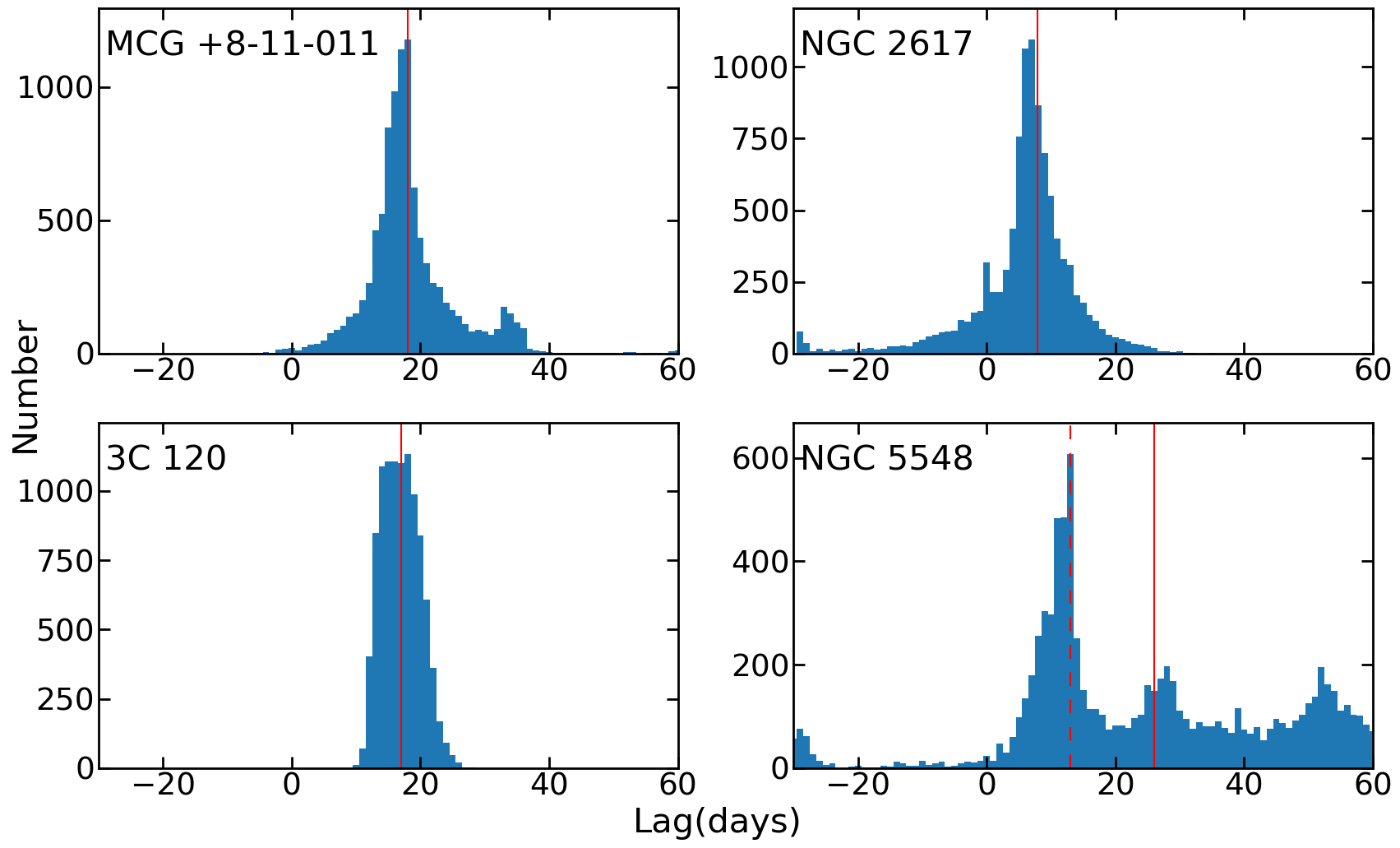}
    \caption{The ICCF-Cut lag distributions with different values of $\alpha$ for 4 Seyfert 1 galaxies. The red line represents the median value of the lag distribution. The H$\alpha$ lags are $18.0_{-4.5}^{+6.5}$ days for MCG +8-11-011, $7.8_{-5.7}^{+4.2}$ days for NGC 2617, $17.0_{-3.1}^{+3.8}$ days for 3C 120 and $26.0_{-16.1}^{+35.2}$ days for NGC 5548 respectively. For NGC 5548, the red dash line represents the peak value of the lag distribution, which is $13.0_{-3.1}^{+48.0}$ days.}
\end{figure*}

We also apply the simulations with the ${\rm H}\beta$ emission line and the inter-continuum lag to other three Seyferts. To investigate the influences of the line lag and cadence, the initial H$\alpha$ lag is set as 10 days for NGC 2617 as the comparison. From Figure 16, we find that for the sources with small lags, like NGC 2617, the dispersion and uncertainty are large. We need higher cadence to obtain the reliable results for AGNs with smaller BLR sizes. From Figure 17 we also find that for NGC 5548, some results of JAVELIN are much smaller than the setting lags. The reason may be because the J index of NGC 5548 (around 1.0) is smaller than other 3 Seyferts (2.0$\sim$3.0), which means that the variability is smaller than other 3 Seyferts. It may explain the small H$\alpha$ lag results of JAVELIN for NGC 5548(panel (f) in Figure 9). It also means that although the dispersion of the simulative lag distributions for JAVELIN is smaller than those of the ICCF-Cut and $\chi^2$ methods, only relying on the result of JAVELIN may have problems. Based on the above simulations, we find that using a single method for the broadband PRM may not be very convincing in some cases. We need to use multiple methods to obtain the time lags. The consistency of the lag distributions from different methods can ensure the reliabilities of the results.

JAVELIN also provides the DPmap model which can be used to calculate the continuum time lag between the continuum and line bands as well as the ${\rm H}\alpha$ line lag. We can compare the H$\alpha$ time lag and continuum lag obtained from the DPmap with the results in Section 4 and the continuum lag in \citet{2018ApJ...854..107F}. The DPmap model assumes that the $r$ band has two components with different time lags. In the MCMC processes, we only request one component to have a $-10\sim10$ day time lag which can be regarded as the inter-continuum lag between the $g$ and $r$ bands. The results for MCG +8-11-011 are shown in Figure 18 as an example. The DPmap model for MCG +8-11-011 shows a similar ${\rm H}\alpha$ time lag distribution at $18.4_{-6.0}^{+2.4}$ days as shown in Figure 6 and Table 2. The ratios between the line and continuum transfer function amplitude given by the DPmap model are close to the real ${\rm H}\alpha$ emission line to the continuum ratio observed in the line band. The continuum lag distribution of $0.5_{-0.3}^{+0.2}$ days is slightly shorter than the inter-continuum lag between the $g$ and $r$ band for ICCF (about 1.7 days) in \citet{2018ApJ...854..107F}. This may be because  the DPmap model includes more parameters than the Pmap model for the same initial data, and the DPmap model is more sensitive to the data quality and tends to yield worse results than the Pmap model. Another reason is that the continuum lags for these local Seyfert 1 AGNs are very small, even smaller than the observational cadence. It needs very high cadence and accuracy to obtain the continuum lag and H$\alpha$ lag simultaneously from the line band. The consistency of ${\rm H}\alpha$ lag results of DPmap model with the results shown in Figure 6 and Table 2 indicates that the influence of the continuum lag in the line band is insignificant and can be ignored for these Seyfert 1 galaxies.

According to the DRW model, the transfer function amplitude (which can be represented by the statistic mean of the ${\rm H}\alpha$ line contribution in the line band) changes very little, but there is a time lag between the ${\rm H}\alpha$ and the continuum, so the contribution of the ${\rm H}\alpha$ in the line  band is not constant. 
Although we use the minimum function in Equation (4) to make sure the calculated ${\rm H}\alpha$ lightcurve contains all the contributions of the ${\rm H}\alpha$ emission line, it still needs to evaluate the change of ${\rm H}\alpha$ ratio for the broadband PRM. According to the ${\rm H}\alpha$ ratio in the line band obtained from the single-epoch spectrum (as shown in Table 1), we change the ${\rm H}\alpha$ ratio from 0.2 to 0.3 with a bin size of 0.01 for MCG +8-11-011, 3C 120 and NGC 5548, and from 0.1 to 0.2 with a bin size of 0.01 for NGC 2617. For each ${\rm H}\alpha$ ratio value, we use the ICCF-Cut to calculate the lag distribution with FR/RSS, then combine all the lag distributions to calculate the median value of the lag with highest posterior density. In Figure 19, MCG +8-11-011, NGC 2617 and 3C 120 show consistent lag distributions with the results presented in Section 4. For NGC 5548, because of the larger uncertainties in the initial photometric data, the median lag is much larger than the results shown in Table 2, but the peak value $13.0_{-3.1}^{+48.0}$ days of the lag distribution for NGC 5548 is very close to the previous results. These consistencies indicate that using the minimum function in Equation (4) is efficient for the H$\alpha$ broadband PRM.

\section{Summary}

By assuming that the continuum flux in the line band equals to a fraction of that in the continuum band, we use the modified method of ICCF (ICCF-Cut) to calculate the ${\rm H}\alpha$ emission line time lags from the lightcurves in the continuum and line broadbands. We also consider the host galaxy contribution to the broadbands and the change of $\alpha$ value for AGNs with longer observational duration to improve the lag results. The lightcurves of extracted H$\alpha$ are similar with the lagged continuum band lightcurves and the lagged simultaneous SRM H$\beta$ lightcurves.

To evaluate the influence of the errors of extracted H$\alpha$ lightcurves and the feasibility of the H$\alpha$ broadband PRM with large uncertainties, we apply the $\chi ^2$ method to weigh of the points in lightcurves by uncertainties. By combining the results of the ICCF-Cut, JAVELIN and $\chi ^2$ methods, we find that the derived ${\rm H}\alpha$ time lags for 4 Seyfert 1 galaxies are consistent with the $R-L$ relationship obtained from the SRM. These AGNs show slightly larger ${\rm H}\alpha$ lags from the broadband PRM than the ${\rm H}\beta$ lags from SRM, which is consistent with previous works and theoretical predictions that the BLR size of the ${\rm H}\alpha$ line is usually larger than that of the ${\rm H}\beta$ emission line.

To confirm our results further, we use the DRW model to simulate the mock lightcurves which have similar parameters with our selected AGNs. By calculating the time lags of these mock lightcurves, we evaluate the reliability of the time lags obtained from the H$\alpha$ broadband PRM and the influences of the ${\rm H}\beta$ line in the continuum band.
By comparing the results of JAVELIN DPmap model with previous results, we find that the continuum lag in the line band can be ignored in broadband PRM for these local Seyfert 1 galaxies whose continuum lags are very small. By calculating the H$\alpha$ lags with different H$\alpha$ ratios, we find that using the minimum function in Equation (4) is efficient for the H$\alpha$ broadband PRM.

From the comparisons of the results from the H$\alpha$ broadband PRM and SRM and the results of simulations, we find that the consistency of the ICCF-Cut, JAVELIN and $\chi^2$ methods can ensure the reliability of the ${\rm H}\alpha$ line lags obtained from the broadband PRM. However, we must admit that all 4 Seyfert 1 galaxies have high quality broadband lightcurves with daily/sub-daily cadences, which enables us to get reliable H$\alpha$ lags. It is difficult to do so for other AGNs with poor quality broadband data. We expect that these broadband PRM methods can be used to study the BLR sizes and BH masses of a large sample of AGNs in the era of large multi-epoch and high cadence photometric sky surveys such as ZTF \citep{2019PASP..131a8003M} and LSST \citep{2017arXiv170804058L} in the near future.

$Acknowledgements$.
We thank the anonymous referee for helpful suggestions. We are thankful for the support of the National Science Foundation of China (11721303, 11927804, and 12133001). We acknowledge the science research grant from the China Manned Space Project with No. CMS-CSST-2021-A06. We acknowledge the support of the staff of the Xinglong 2.16m telescope. This work was partially Supported by the Open Project Program of the CAS Key Laboratory of Optical Astronomy, National Astronomical Observatories, Chinese Academy of Sciences. This work makes use of observations from the Las Cumbres Observatory global telescope network. This work makes use of on observations obtained at the MDM Observatory, operated by Dartmouth College, Columbia University, Ohio State University, Ohio University, and the University of Michigan. The Liverpool Telescope is operated on the island of La Palma by Liverpool John Moores University in the Spanish Observatorio del Roque de los Muchachos of the Instituto de Astrofisica de Canarias with financial support from the UK Science and Technology Facilities Council. This work is based partly on observations obtained with the Apache Point Observatory 3.5 m telescope, which is owned and operated by the Astrophysical Research Consortium. This research has made use of the NASA/IPAC Extra- galactic Database (NED), which is operated by the Jet Propulsion Laboratory, California Institute of Technology, under contract with NASA.

\bibliography{Seyfert.bib}
\bibliographystyle{aasjournal}

\end{document}